\documentclass[12pt,draftcls,journal,onecolumn]{IEEEtran}
\usepackage[nolists,nomarkers,noheads]{endfloat}%figs in the end
\usepackage{amsmath,amssymb,amscd,latexsym,dsfont}
\usepackage{float,color,graphicx,subfigure}
\usepackage{multirow,multicol,array}
\usepackage{comment,cite}
\usepackage{enumerate}
\usepackage{stfloats}
\usepackage{psfrag}
\usepackage{cite}

\title{Towards Fully Optimized BICM Transceivers}
\author{%
\IEEEauthorblockN{Md. Jahangir Hossain\IEEEauthorrefmark{2}, Alex Alvarado\IEEEauthorrefmark{4}, Leszek Szczecinski\IEEEauthorrefmark{3}\\}
\IEEEauthorblockA{
\IEEEauthorrefmark{2}ITR, University of South Australia, Australia\\
\IEEEauthorrefmark{4}Department of Signals and Systems, Communication Systems Group\\ Chalmers University of Technology, Gothenburg, Sweden\\
\IEEEauthorrefmark{3} INRS-EMT, University of Quebec, Montreal,Canada\\
\emph{jahangir.hossain@unisa.edu.au}, \emph{alex.alvarado@chalmers.se}, \emph{leszek@emt.inrs.ca}
}
\thanks{Research supported by the Australian Research Council (ARC) Linkage Project (research grant \#LP 0991370),  NSERC, Canada (research grant \#356807-07), by the Swedish Research Council, Sweden (research grant \#2006-5599), and by the Department of Signals and Systems (Solveig and Karl G Eliasson Memorial Fund), Chalmers University of Technology. Parts of this work have been presented at the International Conference on Communications (ICC) 2010, Cape Town, South Africa, May 2010. Parts of this work are also under review for possible publication at ICC 2011, Kyoto, Japan.}
}%

\newcommand{\tr}[1]{\mathrm{#1}}
\newcommand{\mb}[1]{\boldsymbol{#1}}

\newcommand{\mc}[1]{\mathcal{#1}}

\newcommand{\set}[1]{\{#1\}}
\newcommand{\cd}{\cdot}

\newcommand{\ld}{\ldots}

\newcommand{\ms}[1]{\mathds{#1}}

\newcommand{\ie}{i.e.,~}
\newcommand{\eg}{e.g.,~}

\newcommand{\cf}{cf.~}
\newtheorem{theorem}{Theorem}

\newtheorem{example}{Example}

\begin{document}
\maketitle

\begin{abstract}
Bit-interleaved coded modulation (BICM) transceivers often use equally spaced constellations and a random interleaver. In this paper, we propose a new BICM design, which considers hierarchical (nonequally spaced) constellations, a bit-level multiplexer, and multiple interleavers. It is shown that this new scheme increases the degrees of freedom that can be exploited in order to improve its performance. Analytical bounds on the bit error rate (BER) of the system in terms of the constellation parameters and the multiplexing rules are developed for the additive white Gaussian Noise (AWGN) and Nakagami-$m$ fading channels. These bounds are then used to design the BICM transceiver. Numerical results show that, compared to conventional BICM designs, and for a target BER of $10^{-6}$, gains up to $3$~dB in the AWGN channel are obtained.  For fading channels, the gains depend on the fading parameter, and reach $2$ dB for a target BER of $10^{-7}$ and $m=5$.
\end{abstract}


\section{Introduction}\label{Sec:Introduction}

% BICM
Bit-interleaved coded modulation (BICM) \cite{Zehavi92,Caire98,Fabregas08_Book} is used in most of the existing wireless communication standards, \eg HSPA, IEEE 802.11a/g/n,  DVB, etc. In BICM, the channel encoder and the modulator are separated by a bit-level interleaver which allows the designer to choose the code rate and the constellation independently. BICM maximizes the code diversity, and therefore, outperforms trellis coded modulation (TCM) in fading channels. Compared to TCM, BICM is suboptimal for the additive white Gaussian noise (AWGN) channel because it decreases the minimum Euclidean distance. Nevertheless, its simplicity and flexibility make it an attractive coded modulation scheme even when fading is not present.

%% UEP
BICM appears as a simple out-of-the box coded modulation scheme, however, its full potential is achieved only if its design is optimized. It has been shown in \cite{Alvarado09c} that the interleaver and the code can be jointly designed to exploit the so-called unequal error protection (UEP) caused by the binary labeling of the equally spaced (ES) constellations. In general, if the channel offers UEP to the coded bits, gains in terms of bit error rate (BER) can be obtained, \cf \cite[Sec.~I]{Alvarado09c} and references therein. UEP can also be intentionally introduced when designing the transceiver. For example, UEP can be imposed by allowing unequal power allocation for different bits, by deleting bits using certain patterns (puncturing), by changing the binary labeling of the constellation, or by using signal shaping, \ie by using nonequally spaced (NES) constellations or nonequally likely symbols,  known as \emph{geometrical} and \emph{probabilistic} shaping, respectively.

In this paper, we propose to control the UEP via geometrical shaping. In particular, we use the so-called hierarchical constellations \cite{Vitthaladevuni03}, which have received a great deal of attention in many applications where independent data streams with different qualities must be sent at the same time, e.g.,  in multi-resolution image transmission \cite{Morimoto95,Hossain06b}, and simultaneous voice and multi-class data transmission \cite{Hossain06}. Hierarchical quadrature amplitude modulation (HQAM) constellations are also used in QUALCOMM's MediaFLO \cite{Chari07} and have been standardized for the latest digital video broadcasting-terrestrial (DVB-T2) \cite{ETSI_EN_302_755_v111,ETSI_EN_300_744_v161}. In this paper, we use HQAM constellations in a different context, \ie to transmit only one data  stream with improved error performance. 

Although shaping techniques for BICM have received some attention in the literature (\eg geometrical in \cite{Sommer00,LeGoff03,Barsoum07} and probabilistic in \cite{Fabregas10a}, \cite[Sec.~III-F]{Agrell10b}), they are all based on capacity maximization, which translates into enhanced performance if capacity-approaching codes (turbo or low-density parity-check codes) are used. On the other hand, capacity arguments are less relevant when convolutionally-encoded BICM is considered. In this paper, we study this simple and low-complexity BICM configuration, and thus, we adopt a different approach that aims at the minimization of the BER for a given signal to noise ratio (SNR).

% UEP+M-interleavers
In order to exploit the UEP offered by the constellation, the interleaver must be properly designed. The most commonly interleaver considered in the literature is the single interleaver (S-interleaver) of \cite{Caire98}, which eliminates the UEP caused by the binary labeling\footnote{UEP in BICM was in fact considered an ``undesired feature'' in \cite[Sec.~II]{Caire98}.}. Recently, the the so-called multiple interleavers (M-interleavers) \cite{Alvarado09c} were shown to improve the performance of the system by exploiting the UEP caused by the modulator. In fact, the use of BICM with M-interleavers (BICM-M) corresponds to the original BICM configuration proposed by Zehavi in \cite{Zehavi92}, as well as the original BICM with iterative decoding (BICM-ID) scheme proposed by Li and Ritcey in \cite{Li98}. M-interleavers have also been shown to outperform S-interleavers when BICM-ID is considered \cite{Alvarado10b}.

% MUX
The BICM-M system in \cite{Alvarado09c,Alvarado10b} uses a random bit-level multiplexing (R-MUX) that connects the encoder and the M-interleavers, \ie the M-interleavers assign the coded bits to a particular bit position in the modulator in a pseudo-random fashion (with predetermined probabilities).  By doing this, the dependency of adjacent coded bits is ignored. In this paper, we propose an multiplexing/interleaving inspired by the well-known puncturing strategy based on the periodic elimination of the bits according to a prescribed pattern that matches the temporal structure of the code, \cf \cite{Hagenauer88}.  We show that such a deterministic multiplexing (D-MUX) of the coded bits (followed by random interleaving) notably outperforms the R-MUX used in \cite{Alvarado09c}. 

% Contribution
The contributions of this paper can be summarized as follows. We propose and study a BICM scheme for fading and nonfading channels which considers the use of HQAM constellations (HQAM-BICM), a periodic (and deterministic) bit-level multiplexer, and M-interleavers. It is demonstrated that the degrees of freedom of such a scheme can be exploited to notably improve performance of the system in terms of BER. We use a Gaussian model for the probability density function (PDF) of the L-values passed to the decoder which consider the nonequal spacing between the signal points in  the constellation, and apply it to develop union bounds (UB) on the BER of the system for fading and nonfading channels. These UBs are then used to optimize the transceiver. Presented numerical examples show that the proposed system offers gains over previous BICM configurations (ES-QAM and S-interleavers \cite{Caire98} or ES-QAM and M-interleavers \cite{Alvarado09c}). For the particular cases analyzed in this paper, the gains can be up to $3$~dB for a BER target of $10^{-6}$ in the AWGN channel, and for the Nakagami-$m$ fading channel, the gains can reach 2~dB for a target of $10^{-7}$ and $m=5$.

\section{Proposed BICM Transceiver}\label{Sec:SystemModel}

Throughout this paper, we use boldface letters $\mb{c}_t=[c_{1,t},\ld,c_{N,t}]$ to denote row vectors and capital boldface letters $\mb{C}=[\mb{c}_1^\tr{T},\ld, \mb{c}_M^\tr{T}]^\tr{T}$ to denote a matrix of $M$ rows, where $(\cd)^\tr{T}$ denotes transposition. We denote probability by $\tr{Pr}(\cd)$ and the PDF of a random variable $X$ by $p_X(x)$. A Gaussian distribution with mean value $\mu$ and variance $\sigma^2$ is denoted by $\mc{N}(\mu,\sigma^2)$, the Gaussian PDF with the same parameters by $\psi(\lambda;\mu,\sigma)\triangleq\frac{1}{\sqrt{2\pi}\sigma}\exp(-\frac{(\lambda-\mu)^{2}}{2\sigma^{2}})$, and the Q-function by $Q(x) \triangleq\frac{1}{\sqrt{2\pi}}\int_{x}^{\infty}\exp{\left(-\frac{u^2}{2}\right)}\,du$. The combinations of $i$ nonegative integers such that their sum is $l$ is denoted by $\mathcal{W}_{i}(l)$, where $\mathcal{W}_{i}(l)\triangleq\set{[w_1,\ld,w_{i}]\in(\ms{Z}^+)^{i}:w_1+\ld +w_{i}=l}$.

The HQAM-BICM system model under consideration is shown in Fig.~\ref{Sec:Model:system_model}. In what follows, we describe functionalities of various blocks of such transmission scheme.

\subsection{Encoder, Multiplexing, and Interleaving}\label{EncMuxInt}

The $k_\tr{c}$ vectors of information bits $\mb{i}_{l}=[i_{l,1},\ld,i_{l,N_\tr{c}}]$ with $l=1,\ld,k_\tr{c}$ are encoded by a rate $R=k_\tr{c}/n$ convolutional encoder (ENC) yielding the vectors of coded bits $\mb{c}_p=[c_{p,1},\ld,c_{p,N_\tr{c}}]$ with $p=1,\ld,n$. These are then fed to a deterministic multiplexing (D-MUX) unit which bijectively maps $\mb{C}=[\mb{c}_1^\tr{T}, \ld, \mb{c}_n^\tr{T}]^\tr{T}$ onto $\mb{O}=[\mb{o}_{1}^\tr{T},\ld,\mb{o}_{q}^\tr{T}]^\tr{T}$ with $\mb{o}_{k}=[o_{k,1},\ld,o_{k,N_\tr{s}}]$ and $k=1,\ld,q$. Without loss of generality, we assume $N_\tr{s}q=N_\tr{c}n$. The vector of bits after the D-MUX are fed to $q$ parallel interleavers $\pi_{k}$. The $q$ interleavers are assumed to be independent and give randomly permuted sequences of the bits, \ie $\mb{u}_{k}=\pi_k\set{\mb{o}_{k}}$. Each of the interleavers is connected to the $q$th bit positions in the hierarchical $M$-ary pulse amplitude modulation (HPAM) constellation, where $q=\log_{2}M$.

\begin{figure}[t!]
\psfrag{BICM}[lc][lc][0.65]{\emph{BICM Channel}}
\psfrag{EqCode}[lc][lc][0.65]{\emph{Equiv. Code}}
\psfrag{b1}[cB][cB][0.65]{$\mb{i}_1$}%
\psfrag{ddd}[cc][ct][0.65]{$\vdots$}%
\psfrag{dd}[cc][cc][0.65]{$\ldots$}%
\psfrag{bk}[cB][cB][0.65]{$\mb{i}_{k_\tr{c}}$}%
\psfrag{Enc}[cB][cB][0.65]{ENC}%
\psfrag{c1}[cB][cB][0.65]{$\mb{c}_1$}%
\psfrag{cn}[cB][cB][0.65]{$\mb{c}_n$}%
\psfrag{mux}[cB][cB][0.65][-90]{D-MUX}%
\psfrag{o1}[cB][cB][0.65]{$\mb{o}_{1}$}%
\psfrag{oq}[cB][cB][0.65]{$\mb{o}_{q}$}%
\psfrag{pi1}[cc][cc][0.65]{$\pi_1$}%
\psfrag{piq}[cc][c][0.65]{$\pi_q$}%
\psfrag{u1}[cB][cB][0.65]{$\mb{u}_1$}%
\psfrag{uq}[cB][cB][0.65]{$\mb{u}_q$}%
\psfrag{NES}[cB][cB][0.65]{Hierarch.}%
\psfrag{M2QAM}[cB][cB][0.65]{$M$-PAM}%
\psfrag{Mod}[cB][cB][0.65]{Mapper}%
\psfrag{pi12}[cc][cc][0.65]{$\pi_1^{-1}$}%
\psfrag{pin2}[cc][cc][0.65]{$\pi_q^{-1}$}%
\psfrag{xI}[cB][cB][0.65]{$\mb{x}^\tr{I}$}%
\psfrag{x}[cB][cB][0.65]{$\mb{x}$}%
\psfrag{xQ}[cB][cB][0.65]{$\jmath\mb{x}^Q$}%
\psfrag{h}[cB][cB][0.65]{$\mb{h}$}%
\psfrag{eta}[lB][lB][0.65]{$\mb{z}$}%
\psfrag{yI}[cB][cB][0.65]{$\mb{y}^\tr{I}$}%
\psfrag{yQ}[cB][cB][0.65]{$\mb{y}^Q$}%
\psfrag{y}[cB][cB][0.65]{$\mb{y}$}%
\psfrag{Re}[cB][cB][0.65]{$\Re\set{\cd}$}%
\psfrag{Im}[cB][cB][0.65]{$\Im\set{\cd}$}%
\psfrag{Demod}[cB][cB][0.65]{Demapper}%
\psfrag{U1p}[cB][cB][0.65]{$\tilde{\mb{l}}_{1}$}
\psfrag{Ump}[cB][cB][0.65]{$\tilde{\mb{l}}_{q}$}
\psfrag{L1p}[cB][cB][0.65]{$\mb{l}_1$}
\psfrag{Lnp}[cB][cB][0.65]{$\mb{l}_q$}
\psfrag{demux}[cB][cB][0.65][-90]{DEMUX}%
\psfrag{L1}[cB][cB][0.65]{$\mb{l}_1^\tr{in}$}
\psfrag{Ln}[cB][cB][0.65]{$\mb{l}_n^\tr{in}$}
\psfrag{Dec}[cB][cB][0.65]{DEC}%
\psfrag{b21}[cB][cB][0.65]{$\hat{\mb{i}}_1$}%
\psfrag{b2k}[cB][cB][0.65]{$\hat{\mb{i}}_{k_\tr{c}}$}%
\begin{center}
  \includegraphics[width=\columnwidth]{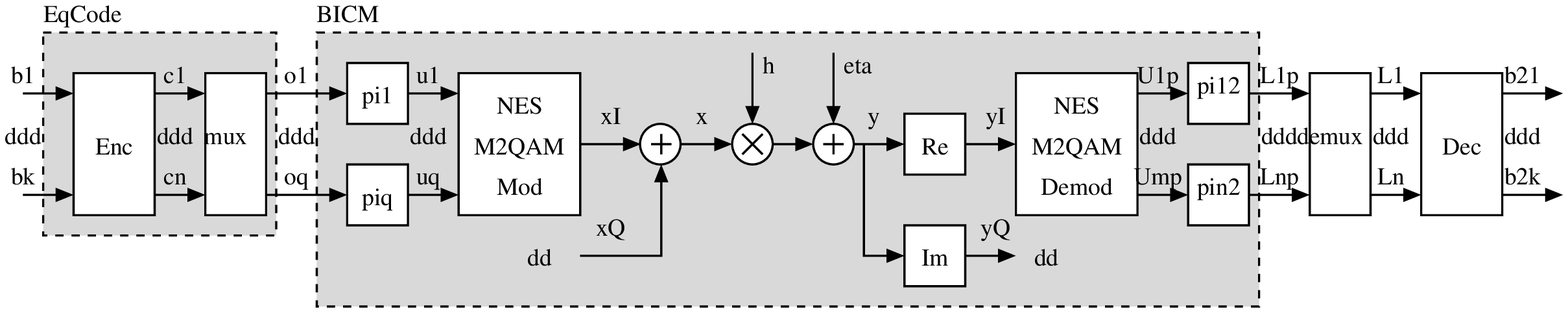}
  \caption{Model of HQAM-BICM transmission:
  a channel encoder followed by the multiplexer (D-MUX), the interleavers ($\pi_1,\ld,\pi_q$), the hierarchical $M$-PAM mapper, the
  channel, and the processing blocks at the receiver's side.}
  \label{Sec:Model:system_model}
\end{center}
\end{figure}

In general, the D-MUX can be defined as a one-to-one mapping between the blocks of $nN_\tr{c}$ and $qN_\tr{s}$ bits, \ie $\set{0,1}^{n N_\tr{c}} \leftrightarrow \set{0,1}^{q N_\tr{s}}$. We define it via an $n\times N_\tr{c}$ matrix $\tilde{\mb{K}}$, whose $(p,t')$th entry is a pair $(k,t)$ where $k\in\set{1,\ld,q}$ and $t\in\set{1,\ld,N_\tr{s}}$. The entry $(k,t)$ indicates that the bit $c_{p,t'}$ is assigned to the $k$th D-MUX's output at time instant $t$, \ie $o_{k,t}=c_{p,t'}$.

The previous definition of the D-MUX is entirely general but difficult to deal with, and thus, in this paper we only consider D-MUX configurations that operate periodically over blocks of $nJ$ bits. We then represent $\tilde{\mb{K}}$ as a concatenation of $N_\tr{c}/J$ matrices $\mb{K}_\tau$, each of dimensions $n\times J$, \ie $\tilde{\mb{K}}=[\mb{K}_0,\ld,\mb{K}_{N_\tr{c}/J-1}]$, where the variable $J$ is called the period of the D-MUX. The entries of $\mb{K}_\tau$ are pairs $(k,t+\tau nJ/q)$ with $k\in \set{1,\ld,q}$ and $t\in\set{1,\ld,nJ/q}$. Without loss of generality, we assume that $(N_\tr{c}\bmod J)=0$ and that $(nJ\bmod q)=0$. To clarify these definitions, consider the following example.
\begin{example}\label{Kexample}
Assume $k_\tr{c}=1$ and $n=2$ ($R=1/2$), $J=3$, and $q=3$ (8-ary constellation). One possible D-MUX is defined by
\begin{align}\label{Kexample_eq}
\mb{K}_\tau=\left[
\begin{array}{ccc}
(1,1+2\tau) & (2,2+2\tau) & (2,1+2\tau)\\
(1,2+2\tau) & (3,2+2\tau) & (3,1+2\tau)
\end{array}\right],
\end{align}
which results in
\begin{align*}
\tilde{\mb{K}}=\left[
\begin{array}{ccc|ccc|c}
(1,1) & (2,2) & (2,1) & (1,3) & (2,4) & (2,3)&\ld\\
(1,2) & (3,2) & (3,1) & (1,4) & (3,4) & (3,3)&\ld
\end{array}\right].
\end{align*}
The mapping between $\mb{C}$ and $\mb{O}$ is then
\begin{align}\label{Ex_C_and_O}
\mb{C}=\left[
\begin{array}{ccc|ccc|c}
c_{1,1}&c_{1,2}&c_{1,3}&c_{1,4}&c_{1,5}&c_{1,6}&\ld\\
c_{2,1}&c_{2,2}&c_{2,3}&c_{2,4}&c_{2,5}&c_{2,6}&\ld
\end{array}\right]
\Longleftrightarrow 
\mb{O}=\left[
\begin{array}{cc|cc|c}
c_{1,1}&c_{2,1} &c_{1,4}&c_{2,4} &\ld\\
c_{1,3}&c_{1,2} &c_{1,6}&c_{1,5} &\ld\\
c_{2,3}&c_{2,2} &c_{2,6}&c_{2,5} &\ld
\end{array}\right].
\end{align}
\end{example}

Since a matrix $\mb{K}_\tau$ is simply a permutation of the set $\set{1,\ld,q}\times\set{1,\ld,nJ/q}$, $(nJ)!$ different matrices $\mb{K}_\tau$ can be generated. However, this number can be reduced since \emph{trivial operations} that do not affect the performance of the system can be applied to $\mb{K}_\tau$. For example, for the matrix in \eqref{Kexample_eq} with $\tau=0$, consider the following two matrices:
\begin{align*}
\mb{K}_0'=\left[
\begin{array}{ccc}
(1,2) & (2,2) & (2,1)\\
(1,1) & (3,1) & (3,2)
\end{array}\right]
\quad
\mb{K}_0''=\left[
\begin{array}{ccc}
(2,1) & (1,1) & (2,2)\\
(3,1) & (1,2) & (3,2)
\end{array}\right].
\end{align*}
The matrix $\mb{K}_0'$ is obtained by permuting the elements of $\mb{K}_0$ such that the first elements in the entries of $\mb{K}_0$ are not altered. Because M-interleavers are used after the D-MUX (\cf Fig.~\ref{Sec:Model:system_model}), the temporal structure of the sequences $\mb{o}_k$ is randomized, and thus, the second elements of the entries are not (which determines to which time bit $c_{k,t}$ is assigned) are not relevant. Consequently, the performance of the system using $\mb{K}_0$ or $\mb{K}_0'$ will be the same. The matrix $\mb{K}_0''$ is obtained by cyclically rotating the columns of the matrix $\mb{K}_0$, which will produce $\tilde{\mb{K}}$ with columns shifted to the left or right. For long coded sequences, \ie $J\ll N_\tr{c}$, the original matrix and its shifted version will yield the same performance.

\subsection{HPAM Constellations}\label{Sec:SystemModel:NESPAM}

In this paper, we consider HQAM constellations labeled by the binary reflected Gray code (BRGC) \cite{Agrell04} presented in \cite{Vitthaladevuni03}. In HQAM constellations, each symbol is a superposition of independently modulated real/imaginary parts, which allows us to focus on the equivalent HPAM constellation, \cf Fig.~\ref{Sec:Model:system_model}. At any time instant $t$, the coded and interleaved bits $[u_{1,t},\ld,u_{q,t}]$ are mapped to an HPAM symbol $x^\tr{I}(t)\in\mc{X}=\set{x_0^\tr{I},\ld,x_{M-1}^\tr{I}}$ using a binary memoryless mapping $\mc{M}:\set{0,1}^q\rightarrow\mc{X}$. Since the mapper is memoryless, from now on we drop the time index $t$. 

\begin{figure}[t!]
\begin{center}
\psfrag{000}[cc][][0.65]{000}
\psfrag{001}[cc][][0.65]{001}
\psfrag{010}[cc][][0.65]{010}
\psfrag{011}[cc][][0.65]{011}
\psfrag{100}[cc][][0.65]{100}
\psfrag{101}[cc][][0.65]{101}
\psfrag{110}[cc][][0.65]{110}
\psfrag{111}[cc][][0.65]{111}
\psfrag{a1}[cB][cB][0.65]{$x_{0}^\tr{I}$}
\psfrag{a2}[cB][cB][0.65]{$x_{1}^\tr{I}$}
\psfrag{a3}[cB][cB][0.65]{$x_{2}^\tr{I}$}
\psfrag{a4}[cB][cB][0.65]{$x_{3}^\tr{I}$}
\psfrag{a5}[cB][cB][0.65]{$x_{4}^\tr{I}$}
\psfrag{a6}[cB][cB][0.65]{$x_{5}^\tr{I}$}
\psfrag{a7}[cB][cB][0.65]{$x_{6}^\tr{I}$}
\psfrag{a8}[cB][cB][0.65]{$x_{7}^\tr{I}$}
\psfrag{2d1}[cc][][0.7]{$2d_{1}$}
\psfrag{2d2}[cc][][0.7]{$2d_{2}$}
\psfrag{2d3}[cc][][0.7]{$2d_{3}$}
\psfrag{k10}[cc][][0.5][0]{$u_1=0$}
\psfrag{k11}[cc][][0.5][0]{$u_1=1$}
\psfrag{k20}[cc][][0.5][0]{$u_2=0$}
\psfrag{k21}[cc][][0.5][0]{$u_2=1$}
\includegraphics[width=0.9\columnwidth]{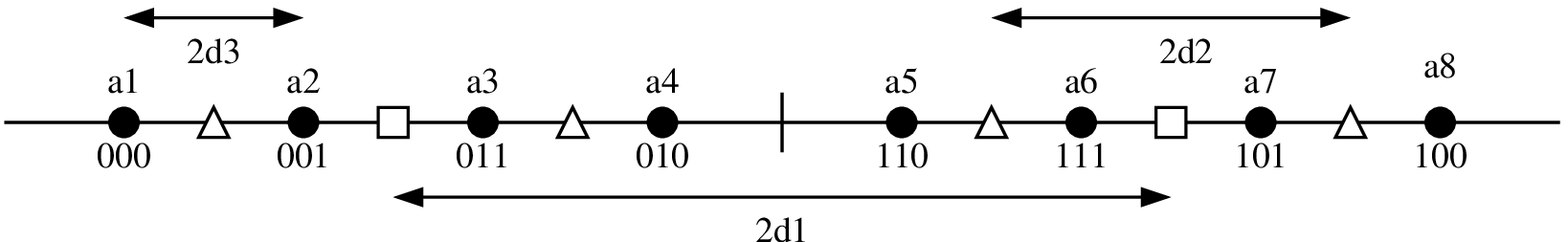}
\caption{HPAM ($M=8$) constellation labeled with the BRGC.}
\label{8PAM}
\end{center}
\end{figure}

We analyze HPAM constellations as the one shown in Fig.~\ref{8PAM} ($M=8$), which are defined by the distances $d_k$ with $k=1,\ld,q$. In this figure, the $M$ constellation points are shown with black circles, where the white squares/triangles are ``virtual'' symbols that help to understand the construction of the HPAM constellation as explained below. We use $k=1,\ld,q$ to denote the bit position of the binary labeling, where $k=1$ represents the left most bit position. The bit value of $k=1$ selects one of the two squares in Fig.~\ref{8PAM}. Similarly, for a given value of the first bit, the bit value for the next position ($k=2$) selects one of the two triangles that surround the previously selected square. Finally, given the bit values for $k=1$ and $k=2$, the bit value of bit position $k=3$ selects one of the two black symbols that surround the previously selected triangle. This selected symbol (black circle) is finally transmitted by the modulator.

We denote the base-2 representation of the integer $0\le j \le M-1$ by the vector $\hat{\mb{b}}(j)=[\hat{b}_1(j), \ld, \hat{b}_{q}(j)]$, where $\hat{b}_{1}(j)$ is the most significant bit of $j$ and $\hat{b}_{q}(j)$ the least significant. This allows us to express the elements $x_j^\tr{I}\in \mc{X}$ of the HPAM constellation as
\begin{align}
x_{j}^\tr{I}=\sum_{k=1}^{q} (-1)^{\hat{b}_k(j)-1}d_k.
\label{symbol:cor}
\end{align}
We also define the normalized constellation parameters as
\begin{align}\label{alphak}
\alpha_{k}\triangleq \frac{d_{k+1}}{d_{1}},
\end{align}
with $k=1,\ld,q-1$. Using \eqref{symbol:cor}, and for equiprobable symbol transmission, the average symbol energy is given by $E_{\tr{s}}=(1+\sum_{k=1}^{q-1} \alpha_{k}^2)d_{1}^2$ with $\alpha_k$ is given by \eqref{alphak}. Throughout this paper, we consider that the constellation is normalized to have unit energy, which translates into the relation $d_1=(\alpha_1^2+\alpha_2^2+\ld+\alpha_{q-1}^2+1)^{-1/2}$.

If the constellation points move freely, it is possible that they cross each other (by having for example $d_3<0$ in Fig.~\ref{8PAM}), and therefore, the binary labeling is not the BRGC anymore. Since in this paper we restrict the analysis to the BRGC, extra constraints on the values of $\alpha_k$ must be added, namely,
\begin{align}
\label{alpha:relation}
\alpha_{k} \geq \sum_{j=k+1}^{q-1} \alpha_j,
\,
\sum_{k=1}^{q-1}\alpha_{k} \leq 1,
\text{ and }
\alpha_{q-1}\geq 0,
\end{align}
where $k=1,\ld,q-1$. The inequalities in \eqref{alpha:relation} are found by solving $(x_{j+1}-x_{j})\geq 0$ with $j=0,\ld,M-2$.

\begin{example}[Constellation parameters for $M=8$]
For $M=8$, the constellation optimization space is formed by two variables, $\alpha_1=d_2/d_1$ and $\alpha_2=d_3/d_1$, \cf \eqref{alphak}. From \eqref{alpha:relation} we have the following constrains
$\alpha_{1} \geq \alpha_2$, 
%\quad
$\alpha_{1}+\alpha_{2} \leq 1$, 
%\quad
and 
$\alpha_2\geq 0$, 
which result in a pair of constellation parameters $(\alpha_1,\alpha_2)$ shown in Fig.~\ref{region:M8}. In this figure, the evolution of the constellation for different values of $(\alpha_1,\alpha_2)$ are shown; the shadowed region represents the values of $(\alpha_1,\alpha_2)$ that give a BRGC-labeled constellation. Particularly important cases are the equally spaced 2-PAM, 4-PAM, and 8-PAM constellations.
\begin{figure}[t!]
\psfrag{2PAM}[lB][lc][0.6]{2-PAM}%
\psfrag{4PAM}[cB][cB][0.6]{4-PAM}%
\psfrag{8PAM}[cB][cc][0.6]{8-PAM}%
\psfrag{xlabel}[cB][cB][0.8]{$\alpha_1$}%
\psfrag{a10}[cB][cB][0.7]{}%
\psfrag{a1025}[cB][cB][0.7]{1/4}%
\psfrag{a105}[cB][cB][0.7]{1/2}%
\psfrag{a1075}[cB][cB][0.7]{3/4}%
\psfrag{a11}[cB][cB][0.7]{1}%
\psfrag{ylabel}[ct][ct][0.8]{$\alpha_2$}%
\psfrag{a20}[rB][rB][0.7]{}%
\psfrag{a205}[rB][rB][0.7]{$1/2$}%
\psfrag{a2025}[rB][rB][0.7]{$1/4$}%
\begin{center}
   \includegraphics[width=0.85\textwidth]{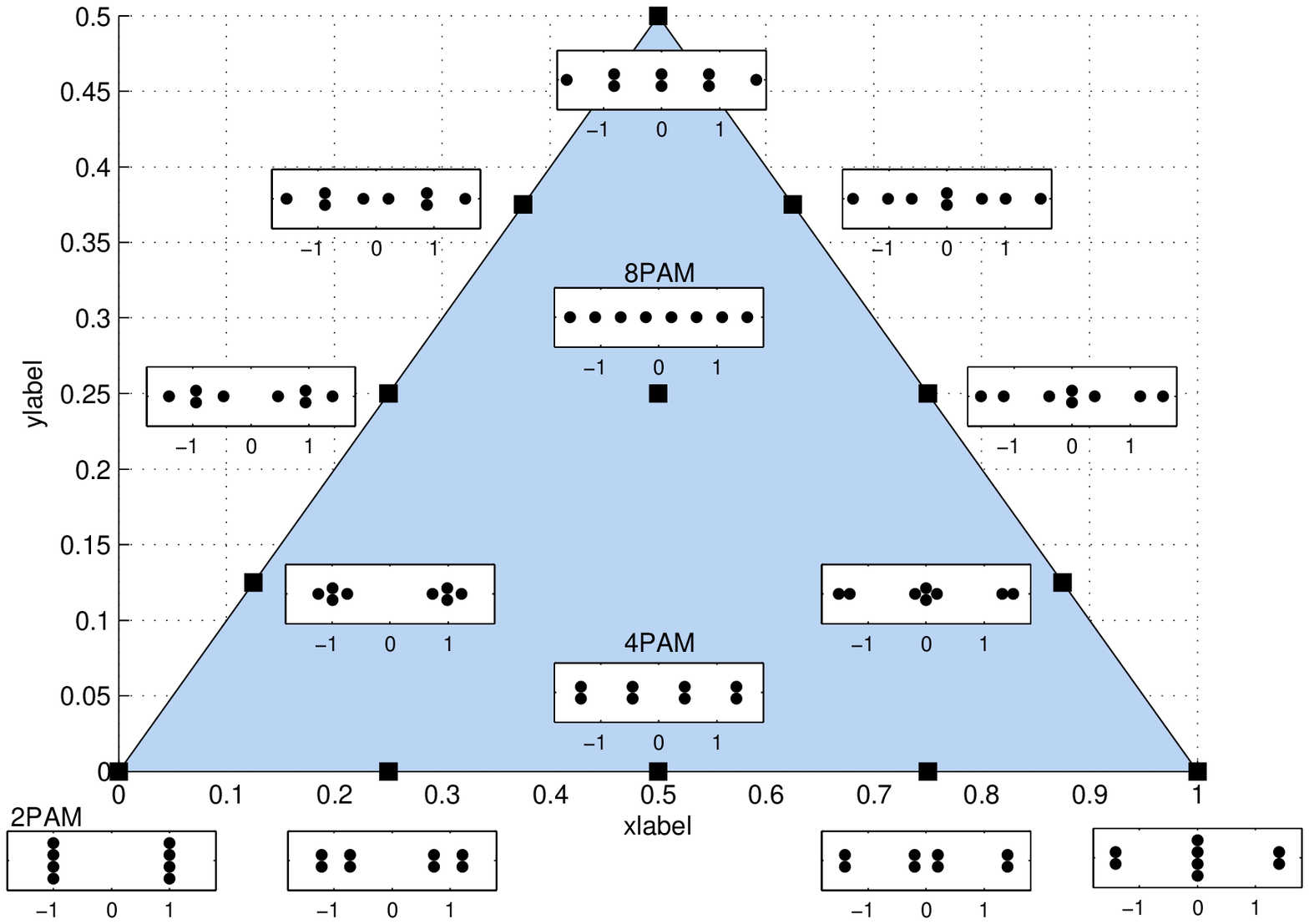}
  \caption{Constellation parameters for an HPAM constellation with $M=8$. The shadowed region shows the values of $(\alpha_1,\alpha_2)$ that give a BRGC-labeled constellation and the 13 filled squared some particular constellations (not at scale). Any point outside this region corresponds to a non-BRGC constellation.}
  \label{region:M8}
\end{center}
\end{figure}
\end{example}

The result of the transmission of a complex symbol $x=x^\tr{I}+\jmath x^Q$ is given by $y=hx+z$, where $h=h^\tr{I}+\jmath h^Q$ is the complex channel gain, and $z$ is a complex Gaussian noise with zero mean and variance $N_0/2$ in each dimension. The amplitude of the channel gain $|h|$ follows a Nakagami-$m$ distribution, and thus, the instantaneous SNR, defined as $\gamma \triangleq \frac{|h|^2}{N_0}$, follows a Gamma distribution, \ie
\begin{equation}
p_{\Gamma}(\gamma;\overline{\gamma})=\frac{\gamma ^{m-1}}{G(m)}\left(\frac{m}{\overline{\gamma}}\right)^m \exp \left(-\frac{m\gamma}{\overline{\gamma}}\right),
\label{gamma:dis}
\end{equation}
where $G(m)$ is the Gamma function, $\overline{\gamma}=\ms{E}_\Gamma[\gamma]$ is the average SNR, and $\Gamma$ is the random variable that represents the instantaneous SNR. The AWGN channel is obtained when $|h|=1$.

At the receiver's side, the real part of the received signal is normalized by the channel gain $h$ ($y^\tr{I}=\Re\set{y/h}$) and passed to the demapper, which computes logarithmic likelihood ratios (L-values) for each bit in the transmitted symbol. The $k$th L-value given the transmitted symbol $x_j$ and the channel gain $h$ (or equivalently, $\gamma$) can be written as \cite{Caire98,Zehavi92,Fabregas08_Book,Alvarado09c}
\begin{align}
\tilde{l}_{k}(y^\tr{I}|x_j,\gamma) 	 = \log\frac{\tr{Pr}(u_k=0|y^\tr{I},\gamma)}{\tr{Pr}(u_k=1|y^\tr{I},\gamma)} 
				 \approx \gamma \biggl[\min_{a \in \mc{X}_{k,1}} \bigl\{(y^\tr{I}-a)^2\bigr\} -\min_{a \in \mc{X}_{k,0}} \bigl\{(y^\tr{I}-a)^2\bigr\}\biggr],
\label{log:max:PDF}
\end{align}
where $\mc{X}_{k,b}$ is the set of symbols labeled with the $k$th bit equal to $b$, and where we have used the so-called max-log approximation \cite{Viterbi98,Caire98,Zehavi92}.

The vectors of L-values calculated by the demapper are then deinterleaved, generating the sequence $\mb{l}_{k}=\pi_k^{-1}(\tilde{\mb{l}}_{k})$ with $k=1,\ld,q$, \cf Fig.~\ref{Sec:Model:system_model}. These L-values are reorganized by the demultiplexer unit (DEMUX), defined as $l_{p,t'}^\tr{in} = l_{k,t}$, which simply inverts the process done by the D-MUX at the transmitter. Finally, these L-values are passed to the the channel decoder which produces an estimate of the transmitted bits. In this paper, we consider convolutional codes and a soft-input Viterbi decoder.

\section{Equivalent Channel Model}\label{Sec:PDFLValuesChannelmodel}

In order to predict the coded BER performance of the system, finding the PDF of the L-values passed to the channel decoder is crucial. In what follows, we develop closed-form expressions for the PDF of L-values for HQAM-BICM transmission as a function of the constellation parameters. These expressions will later be used to compute bounds on the BER of the systems, and then, used to optimize the design of the system. From now on, all the analysis is made for the constituent HPAM constellation (\cf Fig.~\ref{Sec:Model:system_model}), and thus, with a slight abuse of notation, we use $x$ and $y$ to denote the real part of the transmitted symbol and the real part of the received signal, respectively, \ie we skip the superscript $(\cd)^\tr{I}$.

The use of the max-log approximation in \eqref{log:max:PDF} transforms the nonlinear relation between the received signal and the L-values into a piecewise linear relation. Examples of this piece-wise linear relation can be found in literature, see for example \cite[Fig.~3]{Alvarado07d}, \cite[Fig.~2,~Fig.~4]{Benjillali06b}, \cite[Fig.~3]{Hyun05}, or \cite[Table~I]{Raju06}. This piece-wise linear relation has been used to develop expressions for the PDF of the L-values in \eqref{log:max:PDF} using arbitrary signal sets in \cite{Szczecinski08b} (based on an algorithmic approach), closed-form expressions for ES-QAM constellations labeled with the BRGC for the AWGN channel in \cite{Alvarado07d}, and for fading channels in \cite{Szczecinski07f}. Recently, closed-form approximations for the PDF of the L-values for arbitrary signal sets and binary labeling in fading channels have been developed in \cite{Kenarsari10}.

For a given transmitted symbol $x_j$ and SNR $\gamma$, $Y\sim\mc{N}(x_j,1/(2\gamma))$. Using a generalization of the so-called consistent model (CoMod) introduced in \cite{Alvarado07d}, the PDF of the L-values in \eqref{log:max:PDF} can be approximated by
\begin{align}\label{Sec:Preliminaries:PDFLLR:LZCMod}
\tilde{L}_{k}(x_j;\gamma) \sim \mc{N}(\gamma\mu_{k,j},\gamma\sigma_{k,j}^2),
\end{align}
where
 \begin{align}\label{mu_and_sigma}
\mu_{k,j}				= (-1)^{2-\tilde{b}_{k,j}}( \hat{x}_{k,j}-x_{j})^2,\qquad
\sigma_{k,j}^2			= 2( \hat{x}_{k,j}-x_{j})^2=2|\mu_{k,j}|,
\end{align}
$\tilde{\mb{b}}_j=[\tilde{b}_{1,j},\ld,\tilde{b}_{q,j}]$ is the binary label of the symbol $x_j$, and $\hat{x}_{k,j}$ is the  closest symbol to $x_j$ with the opposite bit value at bit position $k$. The model in \eqref{mu_and_sigma} fulfills the consistency condition ($\sigma_{k,j}^2=2|\mu_{k,j}|$), and thus, the Gaussian distribution in \eqref{Sec:Preliminaries:PDFLLR:LZCMod} is completely determined by $\mu_{k,j}$. 

The results in \eqref{Sec:Preliminaries:PDFLLR:LZCMod}--\eqref{mu_and_sigma} can be considered as a particular case of the results in \cite[``Case~1'', Fig.~2, Tab.~1]{Kenarsari10}). In what follows, we will develop generic expressions for $\mu_{k,j}$ in \eqref{mu_and_sigma} in terms of the distances defining the HPAM constellation.

\begin{table}[t!]
\caption{Values of $\mu_{k,j}$ for HPAM with $M=8$ in \eqref{mu_and_sigma}.}
\renewcommand{\arraystretch}{1.2}
\begin{center}
\begin{tabular}{ccccc}
\hline
$x_j$ 	& $\mu_{1,j} $			& $\mu_{2,j} $		& $\mu_{3,j} $\\
\hline

\hline
$x_0$ 	& $+4d_1^2$ 			& $+4d_2^2$ 		& $+4d_3^2$ 			\\
$x_1$ 	& $+4(d_1-d_3)^2$ 		& $+4(d_2-d_3)^2$ 	& $-4d_3^2$	 		\\
$x_2$ 	& $+4(d_1-d_2)^2$ 		& $-4(d_2-d_3)^2$ 	& $-4d_3^2$ 			\\	
$x_3$ 	& $+4(d_1-d_2-d_3)^2$ 	& $-4d_2^2$ 		& $+4d_3^2$ 			\\	
$x_4$ 	& $-4(d_1-d_2-d_3)^2$ 	& $-4d_2^2$ 		& $+4d_3^2$ 			\\	
$x_5$ 	& $-4(d_1-d_2)^2$ 		& $-4(d_2-d_3)^2$ 	& $-4d_3^2$ 			\\	
$x_6$ 	& $-4(d_1-d_3)^2$ 		& $+4(d_2-d_3)^2$ 	& $-4d_3^2$ 			\\	
$x_7$ 	& $-4d_1^2$ 			& $+4d_2^2$ 		& $+4d_3^2$ 			\\	
\hline

\hline
\end{tabular}
\end{center}
\label{8PAM:parameters}
\end{table}

In Table~\ref{8PAM:parameters}, we present the values of $\mu_{k,j}$ in \eqref{mu_and_sigma} as a function of the constellation distances $d_k$ for $M=8$. These values are obtained by direct inspection of Fig.~\ref{8PAM}. From this table we can see that the symmetry of the constellation is reflected in the mean values. For example, for $k=1$, $\mu_{1,3}=-\mu_{1,4}$, $\mu_{1,2}=-\mu_{1,5}$, $\mu_{1,1}=-\mu_{1,6}$, and $\mu_{1,0}=-\mu_{1,7}$. If we analyze the variances, \cf \eqref{mu_and_sigma}, we note that for $k=1$ there are 4 different variances, for $k=2$ two different variances, and for $k=3$ only one. This idea can be generalized, \ie from Table~\ref{8PAM:parameters}, it is possible to infer that for a given $k$, there are $M_k\triangleq \frac{M}{2^k}$ different variances, which are determined by the first $M_k$ values of $j$. This idea was previously used in \cite{Alvarado07d}.

The performance evaluation in Sec.~\ref{Sec:BERPerformance} is based on the transmission of the all-zero sequence, and thus, here we only need to consider positive values of $\mu_{k,j}$ (from \eqref{mu_and_sigma}, $\mu_{k,j}>0$ if $\tilde{b}_{k,j}=0$). From the evolution of $\mu_{k,j}$ in Table~\ref{8PAM:parameters}, we can write a generic closed-form expression for $\mu_{k,j}$ (for any $M$) in terms of constellation distance parameters as follows
\begin{align}\label{mean:k:j}
\mu_{k,j}&= 4\biggl(d_k-\sum_{k'=k+1}^{q} \check{b}_{k'-k}(j)d_{k'}\biggr)^2,
\end{align}
where $j=0,1,\ld,M_k-1$, $k=1,2,\ld, q$, and $\check{\mb{b}}(j)=[\check{b}_q(j), \ld, \check{b}_{1}(j)]$ is the binary representation of the integer $j$ where $\check{b}_{1}(j)$ is the least significant bit.

Using the approximation for the PDF of the L-values given in \eqref{Sec:Preliminaries:PDFLLR:LZCMod}, it is possible to build an equivalent model for the \emph{BICM channel} shown in Fig.~\ref{Sec:Model:system_model} \cite{Alvarado07d,Kenarsari10, Alvarado09a}. This model considers $M_k$ \emph{virtual channels}, each of them determined by $\mu_{k,j}$ in \eqref{mean:k:j}. A given bit $o_{k,t}=0$ can be transmitted through the $j$th virtual channel with a probability given by $\xi_{k,j}=\frac{1}{M_k}$. Then, the PDF of the L-values at the output of the $k$th interleaver $L_{k}$ can be expressed as a Gaussian mixture with density given by
\begin{align}
\label{Gaussain:PDF:mixture}
p_{L_k}(\lambda;\gamma) 	& =\sum_{j=0}^{M_k-1}\xi_{k,j}\psi(\lambda; \gamma\mu_{k,j},2 \gamma\mu_{k,j}) = \frac{2^k}{M}\sum_{j=0}^{M/2^k-1}\psi(\lambda; \gamma\mu_{k,j},2 \gamma\mu_{k,j}).
\end{align}

\section{Performance Analysis}\label{Sec:BERPerformance}

In this section, we develop union bounds (UBs) on the BER of the HQAM-BICM system proposed in Sec.~\ref{Sec:SystemModel} using the PDF of the L-values developed in Sec.~\ref{Sec:PDFLValuesChannelmodel}. 

\subsection{Union Bound}

We define a \emph{remerging sequence} as a path in the trellis of the code (ENC) that leaves the zero state and remerge with it after certain number of trellis stages.  The ENC and the D-MUX are grouped into an ``equivalent code'' (as shown in Fig.~\ref{Sec:Model:system_model}) and characterized by an equivalent weight distribution spectrum (EWDS) $\beta_{\mb{K}}(\mb{w})$ with $\mb{w}=[w_1,\ld,w_q]\in (\mathds{Z}^+)^q$.\footnote{To alleviate the notation, from now on we will refer to the matrix $\mb{K}_\tau$ as $\mb{K}$.} We use the notation $\beta_{\mb{K}}(\mb{w})$ to emphasize that the EWDS depends on the D-MUX configuration determined by $\mb{K}$. This EWDS counts the Hamming weights of all the input sequences that generate remerging sequences with weight $\mb{w}$ at the D-MUX's output.

Using the previous definitions, we can express the (truncated) UB on the BER as
\begin{align}\label{Sec:IntCodeDesign:UB_Ndim}
\tr{BER} \leq \tr{UB}\approx \frac{1}{k_\tr{c}}\sum_{w=w^\tr{free}}^{\hat{w}}\sum_{\mb{w}\in \mc{W}_{q}(w)} \beta_{\mb{K}}(\mb{w}) \tr{PEP}(\mb{w};\overline{\gamma}),
\end{align}
where $\tr{PEP}(\mb{w};\overline{\gamma})$ is the pairwise error probability which represents the probability that the decoder selects a codeword with weight $\mb{w}$ instead of the transmitted all-zero codeword. The PEP can be expressed in terms of the decision variable $D(\mb{w})$ as
\begin{align}\label{PEP_def}
\tr{PEP}(\mb{w};\overline{\gamma}) = \tr{Pr}\set{D(\mb{w})>0},
\end{align}
where
\begin{align}
\label{D_definition}
D(\mb{w}) =\sum_{l=1}^{w_1}L_1^{(l)}+\ld +\sum_{l=1}^{w_q}L_q^{(l)},
\end{align}
and where $L_k^{(i)}$ are independent samples of the random variables representing the L-values whose PDF is given by \eqref{Gaussain:PDF:mixture}. In the following subsections, we will show how to compute $\beta_{\mb{K}}(\mb{w})$ and $\tr{PEP}(\mb{w};\overline{\gamma})$ required to evaluate the UB in \eqref{Sec:IntCodeDesign:UB_Ndim}.

\subsection{Equivalent Weight Distribution Spectrum}\label{Section:Spectrum:Calculation}

The vector $\mb{w}$ corresponds to the Hamming weights of the rows of the matrix $\mb{O}$ generated by remerging sequences represented by the matrix $\mb{C}$. The correspondence between $\mb{C}$ and $\mb{O}$ is determined by the matrix $\mb{K}$ (which defines the D-MUX) as well as by the time at which the remerging sequence starts to diverge. However, due to the periodic structure of $\tilde{\mb{K}}$, only $J$ time instants must be considered. Based on this, the EWDS can be expressed as
\begin{align}\label{beta_K_w2}
\beta_{\mb{K}}(\mb{w}) 	&=\frac{1}{J}\sum_{j=1}^{J}\beta_{\mb{K}}^{(j)}(\mb{w}),
\end{align}
where $\beta_{\mb{K}}^{(j)}(\mb{w})$ represents the EWDS when the decoder starts to diverge at time $t+j$ with arbitrary $t$. We note that the similarities between the EWDS in \eqref{beta_K_w2} and the computation of the WDS of punctured convolutional codes \cite{Hagenauer88}. The following example clarifies the main principle behind \eqref{beta_K_w2}, while more details can be found in \cite[Sec.~II-B]{Hagenauer88}.

\begin{example}[EWDS of the code $(5,7)_8$]\label{CodeK3}
Consider the constraint length $K=3$ convolutional code with polynomial generators $(5,7)_8$, with $w^\tr{free}=5$ and the D-MUX in Example~\ref{Kexample}. For this code, there is one divergent sequence generated by an input sequence with Hamming weight one nd output weight $w^\tr{free}$. The $J=3$ possible input sequences are $\mb{i}_1^{(1)}=[\ld,0,1,0,0,0,0,\ld]$, $\mb{i}_1^{(2)}=[\ld,0,0,1,0,0,0,\ld]$, and $\mb{i}_1^{(3)}=[\ld,0,0,0,1,0,0,\ld]$, which result in the following matrices $\mb{C}$
\begin{align*}
\mb{C}^{(1)}=\left[
\begin{array}{@{~}c@{~}c@{~}c@{~}c@{~}c@{~}c@{~}c@{~}}
\ld&1&0&1&0&0&\ld\\
\ld&1&1&1&0&0&\ld\\
\end{array}\right],
\mb{C}^{(2)}=\left[
\begin{array}{@{~}c@{~}c@{~}c@{~}c@{~}c@{~}c@{~}c@{~}}
\ld&0&1&0&1&0&\ld\\
\ld&0&1&1&1&0&\ld\\
\end{array}\right],
\mb{C}^{(3)}=\left[
\begin{array}{@{~}c@{~}c@{~}c@{~}c@{~}c@{~}c@{~}c@{~}}
\ld&0&0&1&0&1&\ld\\
\ld&0&0&1&1&1&\ld\\
\end{array}\right]
\end{align*}
which by using \eqref{Ex_C_and_O} yield
\begin{align*}
\mb{O}^{(1)}=\left[
\begin{array}{@{~}c@{~}c@{~}c@{~}c@{~}c@{~}c@{~}c@{~}c@{~}}
\ld&1&1&0&\ld \\
\ld&1&0&0&\ld \\
\ld&1&1&0&\ld \\
\end{array}\right],
\mb{O}^{(2)}=\left[
\begin{array}{@{~}c@{~}c@{~}c@{~}c@{~}c@{~}c@{~}c@{~}c@{~}}
\ld&0&0&1&1&\ld \\
\ld&0&1&0&0&\ld \\
\ld&1&1&0&0&\ld \\
\end{array}\right],
\mb{O}^{(3)}=\left[
\begin{array}{@{~}c@{~}c@{~}c@{~}c@{~}c@{~}c@{~}c@{~}c@{~}}
\ld&0&0&0&1&0&\ld \\
\ld&1&0&0&1&0&\ld \\
\ld&1&0&0&1&0&\ld \\
\end{array}\right].
\end{align*}
If we consider only this event at minimum Hamming distance (with input weight one), the final EWDS given by \eqref{beta_K_w2} is obtained by computing the Hamming weights of the rows of $\mb{O}^{(j)}$ with $j=1,2,3$, \ie
\begin{align*}
\beta_{\mb{K}}(\mb{w}) 	&=
\begin{cases}
\frac{2}{3},		& \text{if $\mb{w}=[2,1,2]$}\\
\frac{1}{3},		& \text{if $\mb{w}=[1,2,2]$}
\end{cases}.
\end{align*}
\end{example}

The spectrum $\beta_{\mb{K}}(\mb{w})$ can be numerically calculated using a breadth-first search algorithm \cite{Belzile93}. Clearly, the spectrum must be truncated so that only diverging sequences with total Hamming weight $w_1+\ld+w_q \leq \hat{w}$ are considered, \cf \eqref{Sec:IntCodeDesign:UB_Ndim}.

\subsection{Computation of $\tr{PEP}(\mb{w};\overline{\gamma})$}\label{UB_over_Nakagami_Fading:SMUX}

A common approach for computing the PEP in \eqref{PEP_def} is through the use of the Laplace transform of the PDF of the decision variable (see for example \cite{Kenarsari10} and the references therein).  However, due to its simplicity and accuracy, the saddlepoint approximation (SPA) \cite{Martinez06} has recently attracted considerable interest. In this subsection, we use a generalization of the PEP computation based on the SPA used in \cite{Szczecinski07f,Martinez06,Kenarsari10}, and we apply it to BICM systems based on M-interleavers and HQAM constellations.

Let $\Phi_{L_k}(s; \overline{\gamma})$ be the two-sided Laplace transform of the PDF of the L-value $L_k$ in \eqref{Gaussain:PDF:mixture}, and let $\Phi_{L_k}^{\prime} (s;\overline{\gamma})$ and $\Phi_{L_k}^{\prime \prime} (s;\overline{\gamma})$ be its first and second derivative with respect to $s$, respectively. Let also denote the so-called saddlepoint by $\hat{s}$, where $\hat{s}$ is the solution of $\Phi_{L_k}'(\hat{s};\overline{\gamma})=0$.

\begin{theorem}\label{theo_PEP_SPA}
The PEP in \eqref{PEP_def} can be approximated using the SPA as
\begin{equation}
\tr{PEP}(\mb{w};\overline{\gamma})  \approx   \frac{ 1}{\hat{s}\sqrt{2 \pi}} \left[ \sum_{k=1}^{q}  w_k \frac{\Phi_{L_k}^{\prime\prime}(\hat{s};\overline{\gamma})  }{\Phi_{L_k}(\hat{s};\overline{\gamma})}\right]^{-1/2}  \prod_{k=1}^{q}\left[\Phi_{L_k}(\hat{s};\overline{\gamma}) \right]^{w_k}.
\label{SPA:final}
\end{equation}
\end{theorem}
\begin{IEEEproof}
The proof is given in Appendix~\ref{Appendix.theo_PEP_SPA}.
\end{IEEEproof}

The PEP in Theorem \ref{theo_PEP_SPA} allows us to compute UBs on the BER for the proposed HQAM-BICM for the AWGN and Nakagami-$m$ fading channels, as stated in the following two theorems.

%% Theo 1 (PEP using SPA)
\begin{theorem}\label{theo_UB_SPA_NonFading}
The UB for HQAM-BICM for the AWGN channel ($\gamma=\overline{\gamma}$) using the SPA is
\begin{align}
\tr{UB} &\approx \frac{1}{k_\tr{c}}\sum_{w=w^\tr{free}}^{\hat{w}}\sum_{\mb{w}\in\mathcal{W}_q(w)} \beta_{\mb{K}}(\mathbf{w})\left[ \pi{\gamma} \sum_{k=1}^{q}  w_k \frac{\sum_{j=0}^{M_k-1}\xi_{k,j} \mu_{k,j}\exp\left(-\mu_{k,j}{\gamma}/4\right) }{\sum_{j=0}^{M_k-1}\xi_{k,j}\exp\left(-\mu_{k,j}{\gamma}/4\right)}\right]^{-1/2}\cd \nonumber\\ &\hspace{8cm} \prod_{k=1}^{q}\left[\sum_{j=0}^{M_k-1}\xi_{k,j}\exp\left(-\mu_{k,j}{\gamma}/4\right)\right]^{w_k}. 
\label{UB:SPA:AWGN}
\end{align}
\end{theorem}
\begin{IEEEproof}
The proof is given in Appendix~\ref{Appendix.theo_UB_SPA_NonFading}.
\end{IEEEproof}

\begin{theorem}\label{theo_UB_SPA_Fading}
The UB for HQAM-BICM for the Nakagami-$m$ fading channel using the SPA is
\begin{align}
\tr{UB} &\approx \frac{1}{k_\tr{c}}\sum_{w=w^\tr{free}}^{\hat{w}}\sum_{\mb{w}\in\mathcal{W}_q(w)} \beta_{\mb{K}}(\mathbf{w})  \left[ \pi  \overline{\gamma}  \sum_{k=1}^{q}  w_k \frac{\sum_{j=0}^{M_k-1}\xi_{k,j} \mu_{k,j} \left( \frac{4m}{4m+\overline{\gamma}\mu_{k,j}}\right)^{(m+1)} }{\sum_{j=0}^{M_k-1}\xi_{k,j}\left( \frac{4m}{4m+\overline{\gamma}\mu_{k,j}}\right)^m}\right]^{-1/2}  \nonumber\\ &\hspace{8cm} \prod_{k=1}^{q}\left[\sum_{j=0}^{M_k-1}\xi_{k,j}\left( \frac{4m}{4m+\overline{\gamma}\mu_{k,j}}\right)^m\right]^{w_k}. 
\label{UB:SPA:Fading}
\end{align}
\end{theorem}
\begin{IEEEproof}
The proof is given in Appendix~\ref{Appendix.theo_UB_SPA_Fading}.
\end{IEEEproof}

The expressions in Theorems~\ref{theo_UB_SPA_NonFading} and \ref{theo_UB_SPA_Fading} explicitly show the mean values $\mu_{k,j}$ (which depend on the constellation parameters) and the EWDS of the code, and thus, they can be used to optimize the performance of the system. Moreover, we recognize that the PEP computation in \eqref{PEP_def} for both AWGN and Nakagami-$m$ fading channels can be done directly using the PDF of the L-values in \eqref{Gaussain:PDF:mixture} (\cf \cite{Alvarado09c} for the AWGN channel). Nevertheless, we used the SPA because it results in ready-to-use formulas, \cf \eqref{UB:SPA:AWGN} and \eqref{UB:SPA:Fading}.
 
We conclude this section by noting that the BICM models in \cite{Alvarado09c} and \cite{Caire98} can be regarded as a particular cases of the model we introduced in this paper. The model in \cite{Alvarado09c} can be obtained using our model if $J\rightarrow N_\tr{c}$ and the first elements of the entries $(k,t)$ of $\tilde{\mb{K}}$, which represent the assignement of the coded bits to a particular interleaver, are randomly selected with predetermined probabilities. The BICM with S-interleavers (BICM-S) configuration of \cite{Caire98} can be obtained by letting $J\rightarrow N_\tr{c}$ and by selecting a matrix $\tilde{\mb{K}}$ with elements randomly permuted. By doing this, we assure that the coded bits are uniformly assigned over time and also over the bit positions.

\section{Numerical Results}\label{Sec:NumericalResults}

In this section, we present numerical examples that illustrate the gains that can be obtained by using an optimized HQAM-BICM system. In particular, we analyze two practically relevant spectral efficiencies: 1~bit/dimension and 1.5~bit/dimension. We use a rate $R=1/2$ optimum distance spectrum convolutional code with constraint length $K=3$ and generator polynomial $(5,7)_8$. The decoding is based on the soft-input Viterbi algorithm without memory truncation, and the block length used for simulation is $N_\tr{c}=24000$. The optimization of the UB was carried out numerically via an exhaustive search over the valid range of constellation parameter(s) with a step size of $0.01$.  In the following subsections, we use the names 2-PAM, 4-PAM, and 8-PAM to refer to the ES-PAM constellations.  

%% 1~bit/dimension
\subsection{Spectral efficiency 1~bit/dimension}

For this particular case ($n=q=2$) there is only one constellation parameter, \ie $\alpha_1$. From Fig.~\ref{region:M8} (with $\alpha_2=0$), we observe three cases of particular interest: $\alpha_1=0$ gives a 2-PAM constellation, $\alpha_1=1/2$ gives a 4-PAM constellation, and $\alpha_1=1$ gives a three-point constellation. We consider a D-MUX with period $J=2$ which result in only four different matrices:
\begin{align*}
\mb{K}^{(1)}=\left[
\begin{array}{cc}
(1,1) & (1,2) \\
(2,1) & (2,2) 
\end{array}\right],
\mb{K}^{(2)}=\left[
\begin{array}{cc}
(1,1) & (2,1)\\
(1,2) & (2,2) 
\end{array}\right],
\end{align*}
\begin{align*}
\mb{K}^{(3)}=\left[
\begin{array}{cc}
(2,1) & (1,1)\\
(1,2) & (2,2) 
\end{array}\right],  \mbox{and}~
\mb{K}^{(4)}=\left[
\begin{array}{cc}
(2,1) & (2,2)\\
(1,1) & (1,2) 
\end{array}\right].
\end{align*}
 For this specific example, the D-MUX configurations given by $\mb{K}^{(1)}$, and $\mb{K}^{(4)}$ are identical to the two possible R-MUX configurations \cite{Alvarado09c}, \ie when all the coded bits from one encoder's output are assigned to one modulator's input.

\begin{figure}[t!]
\psfrag{AWGN}[cB][cB][0.8]{AWGN Channel}%
\psfrag{Fading}[cB][cB][0.8]{Nakagami-$m$ Fading Channel}%
\psfrag{ylabel}[ct][ct][0.8]{$\tr{UB}$}%
\psfrag{xlabel}[cB][cB][0.8]{$\alpha_1$}%
\psfrag{K1212}[lc][lc][0.75]{$\mb{K}=\mb{K}^{(2)}$}%
\psfrag{K1221}[lc][lc][0.75]{$\mb{K}=\mb{K}^{(3)}$}%
\psfrag{K1122}[lc][lc][0.75]{$\mb{K}=\mb{K}^{(1)}$}%
\psfrag{K2211}[lc][lc][0.75]{$\mb{K}=\mb{K}^{(4)}$}%
\psfrag{SNR-10dB}[lc][lc][0.8]{$\overline{\gamma}=10~\tr{dB}$}%
\psfrag{m-1}[lc][lc][0.8]{$m=1$}%
\psfrag{SNR-16dB}[lc][lc][0.8]{$\overline{\gamma}=16~\tr{dB}$}%
\psfrag{m-5}[lc][lc][0.8]{$m=5$}%
\psfrag{SNR-12dB}[lc][lc][0.8]{$\overline{\gamma}=12~\tr{dB}$}%
\begin{center}
  \includegraphics[width=1\columnwidth]{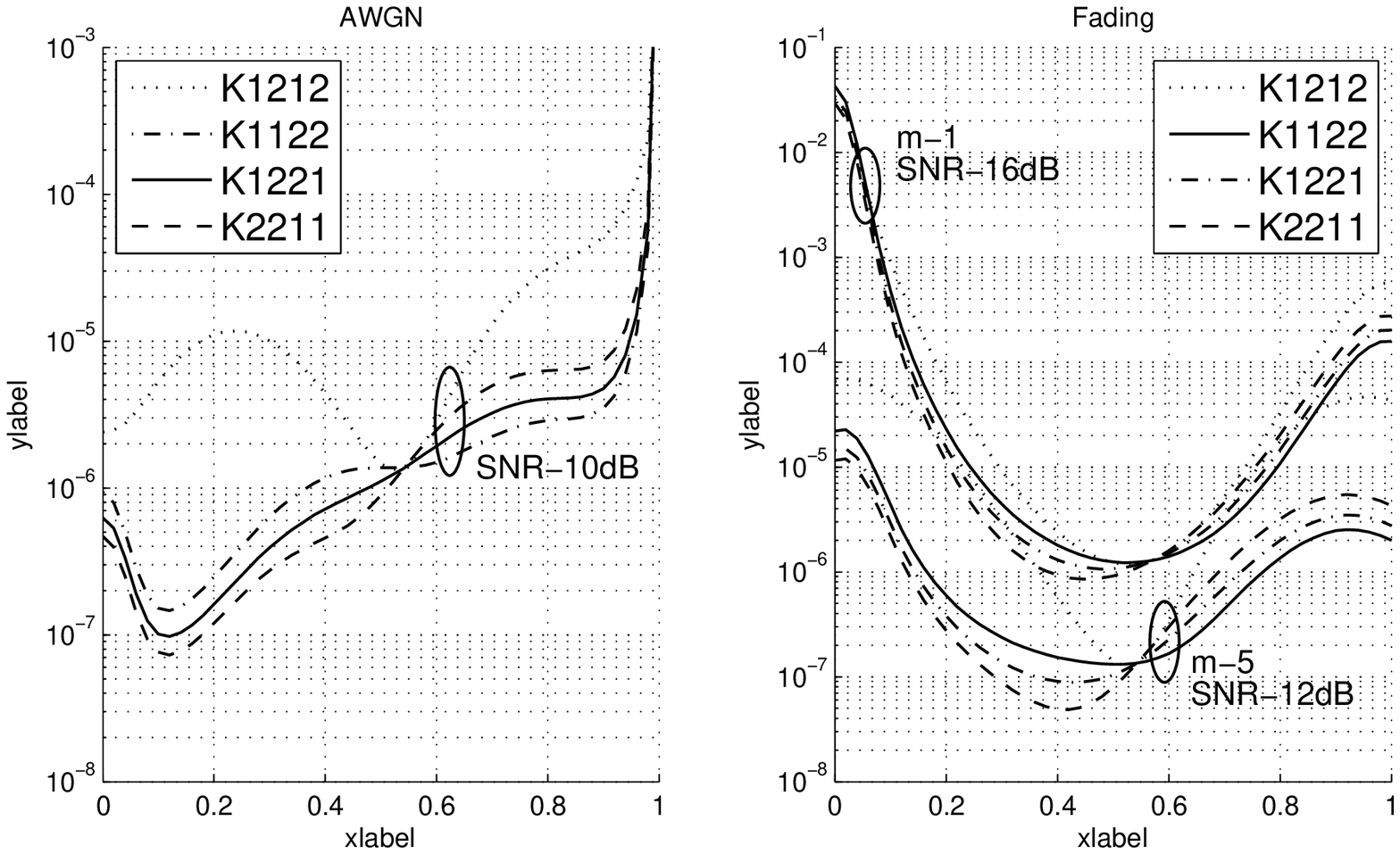}
  \caption{UB versus the constellation parameter $\alpha_{1}$ for the AWGN channel (left) and for Nakagami-$m$ fading channels (right) for different D-MUX configurations and different values of $\overline{\gamma}$. The UBs are given by Theorem~\ref{theo_UB_SPA_NonFading} and Theorem~\ref{theo_UB_SPA_Fading}, respectively.}
  \label{UBvsAlpha_AWGN_and_fading}
\end{center}
\end{figure}

For both the AWGN channel and Nakagami-$m$ fading channels, we first study the behavior of the UBs given by Theorem~\ref{theo_UB_SPA_NonFading} and Theorem~\ref{theo_UB_SPA_Fading}, respectively\footnote{All the results presented in this subsection were obtained using $\hat{w}=125$ for the AWGN channel and $\hat{w}=30$ for Nakagami-$m$ fading channels, \cf~\eqref{Sec:IntCodeDesign:UB_Ndim}.}, as a function of the constellation parameter $\alpha_1$. The UB is shown in Fig.~\ref{UBvsAlpha_AWGN_and_fading} for the four different D-MUX configurations and different values of $\overline{\gamma}$. This figure shows that for a given constellation parameter, different D-MUX configurations give different BER performances. In particular, for all the 3 cases in Fig.~\ref{UBvsAlpha_AWGN_and_fading}, when $\alpha_1=1/2$ (4-PAM) is considered, the lowest UB performance is obtained $\mb{K}=\mb{K}^{(4)}$ (but this changes if another value of $\alpha$ is chosen). This is equivalent to the R-MUX when all the bits from $(7)_8$ are assigned to $k=1$ and the coded bits from $(5)_8$ to $k=2$, which was shown in \cite{Alvarado09c} (only for the AWGN case). From this figure, it is also clear that  the 4-PAM constellation is  suboptimal, i.e., by selecting another value of $\alpha_1$, gains are obtained. More particularly, for the AWGN case, by changing the value of $\alpha_1$ from $\alpha_1=1/2$ to $\alpha_1=0.12$, the UB decreases from $\tr{UB}\approx 0.8\cd10^{-6}$ to $\tr{UB}\approx 0.7\cd10^{-7}$. The gains for the Nakagami-$m$ fading channel are smaller but still visible.
    
The UB depends on the average SNR, the constellation parameter, the D-MUX configuration, and $m$ (in fading channels). Therefore, to obtain the optimal design, one needs to jointly optimize $\mb{K}$ and $\alpha_1$ for each value of $\overline{\gamma}$ and $m$ (in fading channels) that minimize the UB. Because of the nature of the UB, the bound is tight only for BER below certain value (typically $10^{-3}$ or $10^{-4}$), and thus, the optimization will be valid only for BER below this limit. From now on, we denote the values that minimize the UB for a given $\overline{\gamma}$ by $\alpha_1^*(\overline{\gamma})$ and $\mb{K}^*(\overline{\gamma})$.

\begin{figure}[t!]
\psfrag{ylabel1}[ct][ct][0.8]{$\alpha_1^*(\overline{\gamma})$}%
\psfrag{xlabel1}[cB][cB][0.8]{$\overline{\gamma}$~[dB]}%
\psfrag{ylabel2}[ct][ct][0.8]{$\tr{UB}$}%
\psfrag{xlabel2}[cB][cB][0.8]{$\alpha_1$}%
\psfrag{m-1}[lc][lc][0.8]{$m=1$}%
\psfrag{m-2}[lc][lc][0.8]{$m=2$}%
\psfrag{m-5}[lc][lc][0.8]{$m=5$}%
\psfrag{m-7}[lc][lc][0.8]{$m=7$}%
\psfrag{m-20}[lc][lc][0.8]{$m=20$}%
\psfrag{m-50}[lc][lc][0.8]{$m=50$}%
\psfrag{AWGN}[lc][lc][0.8]{AWGN}%
\begin{center}
 \includegraphics[width=1\columnwidth]{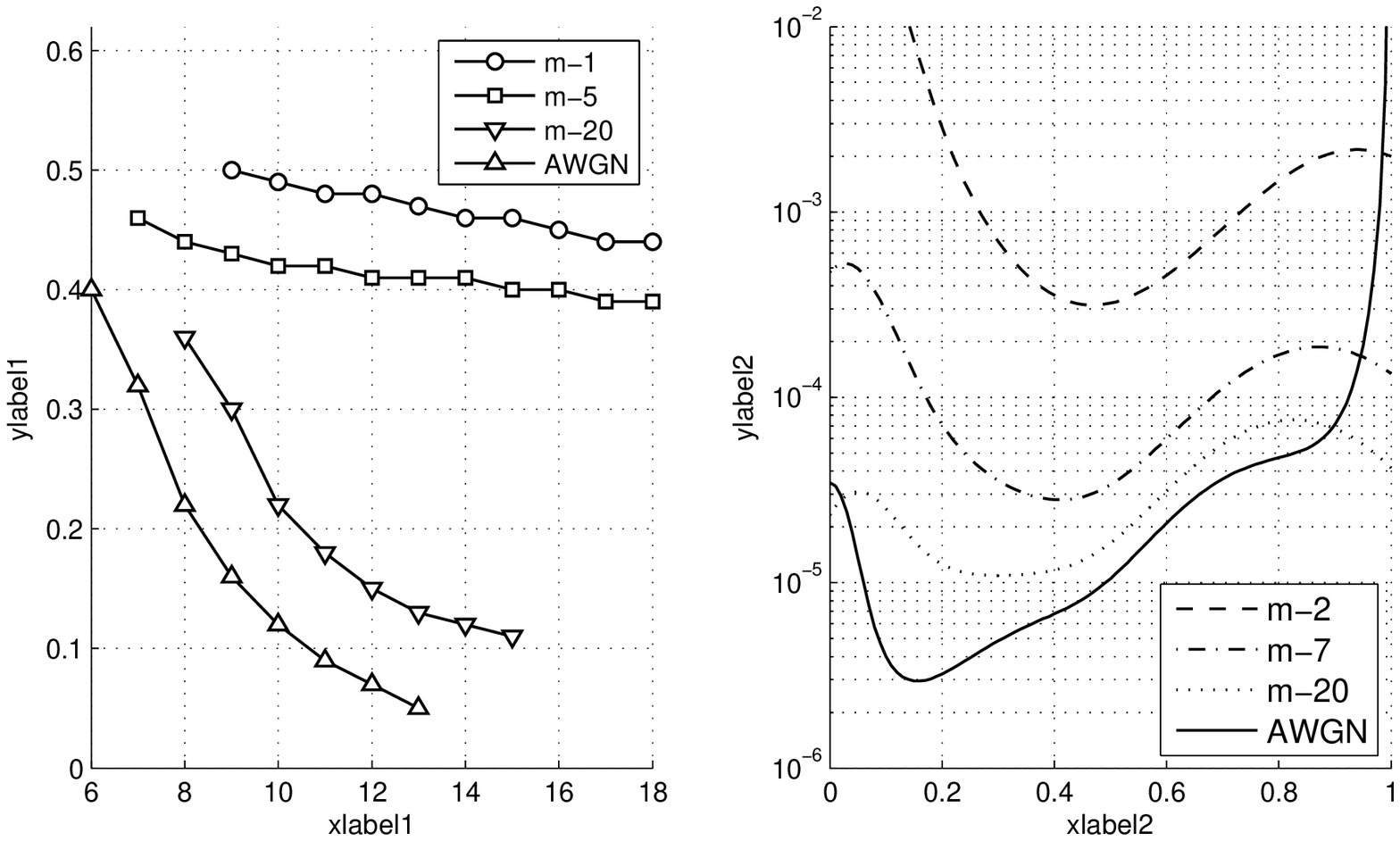}
  \caption{The optimal constellation parameter $\alpha_1^*(\overline{\gamma})$ versus average SNR $\overline{\gamma}$ for $m=1$, $m=5$, $m=20$, and for the AWGN channel (left) and the UB versus $\alpha_1$ for $\overline{\gamma}=9~\tr{dB}$ and $m=2$, $m=7$, $m=20$ and the AWGN channel (right). The UBs are given by Theorem~\ref{theo_UB_SPA_NonFading} and Theorem~\ref{theo_UB_SPA_Fading}, respectively.}
  \label{Alpha_vs_SNR_and_UB_vs_alpha_many}
\end{center}
\end{figure}

We performed a numerical optimization over $\mb{K}$ and $\alpha_1$ for $m=1$, $m=5$, $m=20$, and for the AWGN channel for the values of $\overline{\gamma}$ that give a BER of interest ($\tr{BER}\leq 10^{-3}$). The results obtained showed that in such a case, the optimum D-MUX is always given by $\mb{K}^*(\overline{\gamma})=\mb{K}^{(4)}$. The values of $\alpha_1^*(\overline{\gamma})$ obtained in the optimization are shown in Fig.~\ref{Alpha_vs_SNR_and_UB_vs_alpha_many} (left). This results show that when the fading is severe, the optimal constellation is close to a 4-PAM constellation, and in fact does not depend much on the value of $\overline{\gamma}$. On the other hand, for the AWGN channel, the dependency on the average SNR is notable and a 4-PAM constellation is far from the optimum. In Fig.~\ref{Alpha_vs_SNR_and_UB_vs_alpha_many} (right), we show the behavior of the UB as a function of $\alpha_1$ for a given average SNR $\overline{\gamma}=9~\tr{dB}$ and different channel conditions. This figure shows how the value $\alpha_1^*(\overline{\gamma})$ evolves from $\alpha_1^*(\overline{\gamma})\approx 0.5$ (for $m=2$) to $\alpha_1^*(\overline{\gamma})\approx 0.16$ for the AWGN case.

\begin{figure}[t!]
\psfrag{ylabel}[ct][ct][0.8]{$\tr{BER}$}%
\psfrag{xlabel}[cB][cB][0.8]{$\overline{\gamma}$~[dB]}%
\psfrag{BICM-S-ES-QAM}[lc][lc][0.8]{Traditional BICM-S (ES-QAM) \cite{Caire98}}%
\psfrag{R-MUX-optimum-ES-QAM}[lc][lc][0.8]{R-MUX (ES-QAM) \cite{Alvarado09c}}%
\psfrag{S-MUX-optimum-NES-QAM}[lc][lc][0.8]{Optimal HQAM-BICM}%
\psfrag{Sim-BICM-S}[lc][lc][0.8]{Sim. Traditional BICM-S (ES-QAM)}%
\psfrag{Sim-RMUX}[lc][lc][0.8]{Sim. R-MUX (ES-QAM)}%
\psfrag{Sim-SMUX}[lc][lc][0.8]{Sim. Optimal HQAM-BICM}%
\psfrag{m-1}[lc][lc][0.8]{$m=1$}%
\psfrag{m-5}[lc][lc][0.8]{$m=5$}%
\psfrag{m-inf}[rc][rc][0.8]{AWGN}%
\begin{center}
 \includegraphics[width=0.9\columnwidth]{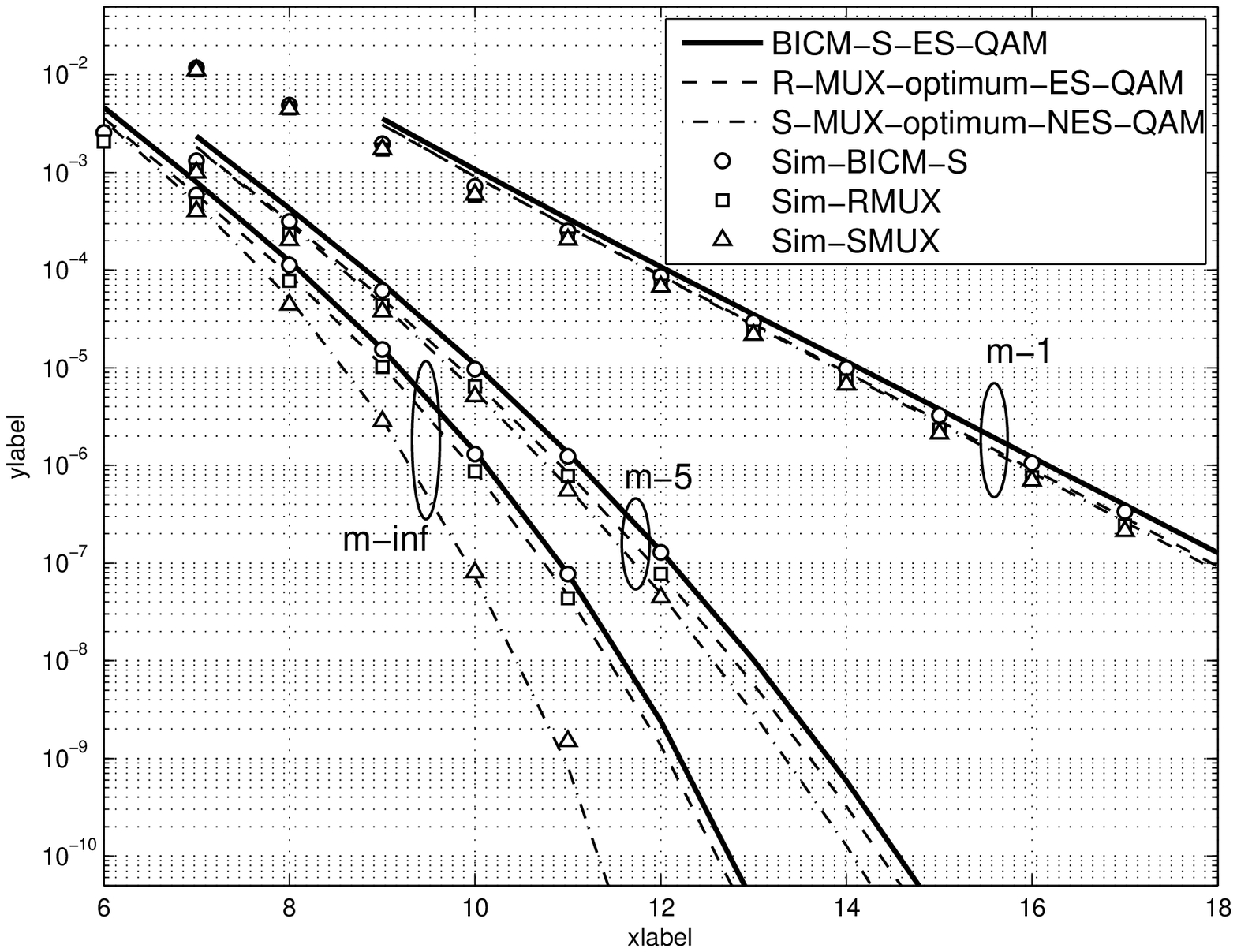}
  \caption{UBs (lines) for $n=q=2$ (1~bit/dimension), for Nakagami-$m$ fading channels and the AWGN channel given by Theorem~\ref{theo_UB_SPA_NonFading} and Theorem~\ref{theo_UB_SPA_Fading}, respectively. Numerical simulations are also included (markers). The proposed HQAM-BICM system uses the optimal $\mb{K}^*(\overline{\gamma})=\mb{K}^{(4)}$ and $\alpha_{1}^*(\overline{\gamma})$ shown in Fig.~\ref{Alpha_vs_SNR_and_UB_vs_alpha_many} (left). The BICM-S system of \cite{Caire98} with a 4-PAM constellation and the BICM system with R-MUX and a 4-PAM constellation of \cite{Alvarado09c} are shown for comparison.}
  \label{BERvsSNR_fading_opt_alpha_opt_MUX_m_1_m_5_m_inf_v2_with_sim}
\end{center}
\end{figure}

We conclude this subsection by presenting the BER performance obtained by using the proposed HQAM-BICM system, where the constellation and the D-MUX are optimized for each SNR. The results are presented in Fig.~\ref{BERvsSNR_fading_opt_alpha_opt_MUX_m_1_m_5_m_inf_v2_with_sim}, where we also show the results obtained by the conventional BICM-S system of \cite{Caire98}, and the one studied in \cite{Alvarado09c} (R-MUX), both of them using a 4-PAM constellation. The results in this figure confirm the tightness of the UBs developed in this paper (for $\tr{BER}\leq 10^{-3}$)\footnote{We note a slight mismatch between the simulations and the bounds for $m=1$ (for the three configurations shown in Fig.~\ref{BERvsSNR_fading_opt_alpha_opt_MUX_m_1_m_5_m_inf_v2_with_sim}). We conjecture this is caused by the approximation used to model the PDF of the L-values, \cf Sec.~\ref{Sec:PDFLValuesChannelmodel}. A similar mismatch is observed in Fig.~\ref{BERvsSNR:Example2:Fading}, \cf Sec.~\ref{1_5}.}. They also confirm that a joint optimization of the D-MUX and the constellation outperforms the previous designs. More particularly, for a BER target of $10^{-7}$, the obtained gains are approximately 1~dB for the AWGN channel and 0.4~dB for $m=5$. These gains decrease when the fading in the channel increases, and for $m=1$ (Rayleigh fading channel), they are marginal.

%% 1.5~bit/dimension
\subsection{Spectral efficiency 1.5~bit/dimension}\label{1_5}
We consider 8-ary constellations ($q=3$), which together with the rate $R=1/2$ code ($n=2$) gives a spectral efficiency of $1.5$~bits/dimension. In this case, the constellation is defined by the pair $(\alpha_1,\alpha_2)$. From Fig.~\ref{region:M8}, we see that the 8-PAM constellation is obtained with $\alpha_2=1/4$ and $\alpha_1=1/2$ and that 4-ary constellations are obtained with $\alpha_2=0$ (4-PAM with $\alpha_1=1/2$). We consider D-MUX configurations with the shortest possible period, \ie $J=3$, for which there will be a total of thirty different D-MUX configurations. The UBs presented in this subsection are evaluated for $\hat{w}=30$ for both fading and the AWGN channel.

For the AWGN channel, and an average SNR $\overline{\gamma}~[\tr{dB}]\in\set{10, 11,\ld,15}$ (which give a UB below $10^{-3}$), we obtained the optimal D-MUX configuration $\mb{K}^{\ast}(\overline{\gamma})$ and constellation parameters ($\alpha_1^{\ast}(\overline{\gamma}),\alpha_2^{\ast}(\overline{\gamma}))$ using Theorem~\ref{theo_UB_SPA_NonFading} and an exhaustive search. The optimal matrix for all $\overline{\gamma}~[\tr{dB}]\in\set{10, 11,\ld,15}$ was found to be
\begin{align}\label{optimal_K_n_2_q_3}
\mb{K}^{\ast}(\overline{\gamma}) 	= \begin{bmatrix} (1,1) & (2,1) & (3,1)\\ (3,2) & (2,2) & (1,2) \end{bmatrix}
\end{align}
and the optimal constellation parameters are
\begin{align}\label{optimal_a1_a2_n_2_q_3}
(\boldsymbol{\alpha}_1^{\ast}(\overline{\gamma}), \boldsymbol{\alpha}_2^{\ast}(\overline{\gamma}))=
[(0.46,0)\, (0.45,0)\, (0.44,0)\, (0.43,0) \, (0.43,0)\, (0.43,0)]
\end{align}
The results in \eqref{optimal_a1_a2_n_2_q_3} indicate that the optimal constellation is a 4-ary constellation ($\alpha_{2}=0$) with $\alpha_1 \approx 0.45$, which translates into a system where the third output of the D-MUX (\cf \eqref{optimal_K_n_2_q_3}) is completely eliminated. 

Another way of interpreting the results in \eqref{optimal_K_n_2_q_3}--\eqref{optimal_a1_a2_n_2_q_3} is that for this code, the minimum BER is obtained when the original rate 1/2 code is punctured (giving a rate $R=3/4$) and transmitted with a 4-ary constellation (with $\alpha_1 \approx 0.45$), and a puncturing pattern given by $\mb{P} \triangleq \bigl[ [1,0]^\tr{T}, [1,1]^\tr{T},[0,1]^\tr{T}\bigr]$,
%\begin{align}\label{puncturing}
%\mat{A} 		= \begin{bmatrix} 1 & 1 & 0\\ 0 & 1 & 1 \end{bmatrix},
%\end{align}
where following the notation of \cite{Hagenauer88}, the columns of $\mb{P}$ have a meaning of time and a $0$ denotes a puncture. 

An intuitive explanation of the previous results is the following. In a coded modulation system, and for high SNR values, there is a trade-off between the minimum Euclidian distance of the constellation and the minimum hamming distance of the code. By puncturing this code, its minimum Hamming distance $w^\tr{free}=5$ will decrease. On the  other hand, by reducing the constellation size (from 8-ary to 4-ary), the minimum Euclidian distance increases. For this particular code, the improvement due to an increased minimum Euclidian distance is larger than the degradation due to a decrease minimum Hamming distance, and thus, the optimal solution is given by \eqref{optimal_K_n_2_q_3}--\eqref{optimal_a1_a2_n_2_q_3}.

\begin{figure}[t!]
\psfrag{ylabel}[ct][ct][0.8]{$\tr{BER}$}%
\psfrag{xlabel}[cB][cB][0.8]{$\overline{\gamma}$~[dB]}%
\psfrag{BICM-S-ES-QAM}[lc][lc][0.8]{Traditional BICM-S (ES-QAM) \cite{Caire98}}%
\psfrag{R-MUX-optimum-ES-QAM}[lc][lc][0.8]{R-MUX (ES-QAM) \cite{Alvarado09c}}%
\psfrag{R-MUX-optimum-NES-QAM}[lc][lc][0.8]{R-MUX with HQAM}%
\psfrag{punctured}[lc][lc][0.8]{Best punctured BICM-S (ES-QAM)}%
\psfrag{S-MUX-optimum-NES-QAM}[lc][lc][0.8]{Optimal HQAM-BICM}%
\psfrag{Simulations}[lc][lc][0.8]{Simulations}%
\begin{center}
 \includegraphics[width=0.9
 \columnwidth]{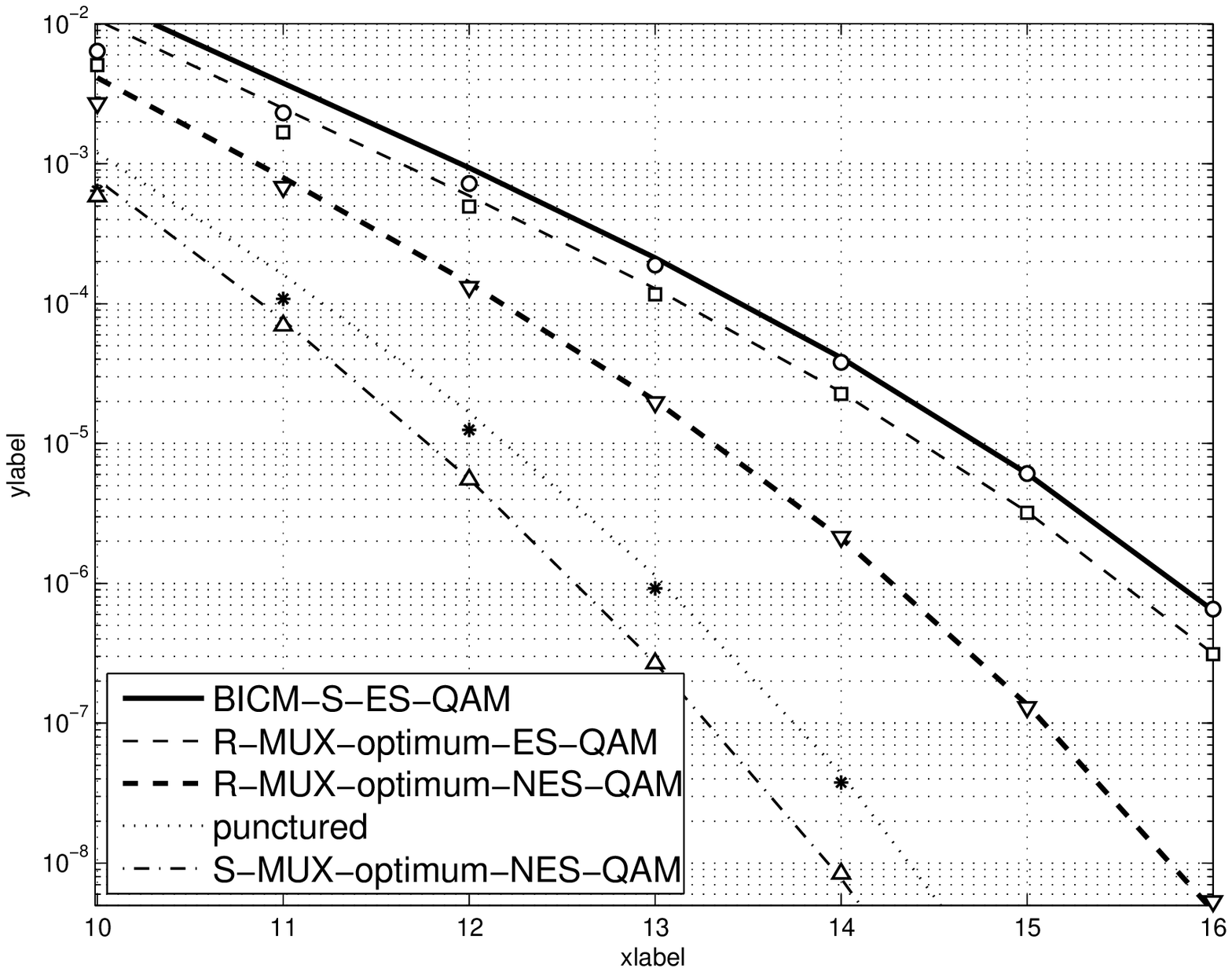}
  \caption{UB (lines) for $n=2$ and $q=3$ (1.5~bit/dimension) for the AWGN channel given by Theorem~\ref{theo_UB_SPA_NonFading}. The simulation results are shown with markers. The proposed HQAM-BICM system uses the optimized parameters in \eqref{optimal_K_n_2_q_3}--\eqref{optimal_a1_a2_n_2_q_3}. The BICM-S system of \cite{Caire98} with an 8-PAM constellation and the BICM system with R-MUX and an 8-PAM constellation of \cite{Alvarado09c} are shown for comparison. The R-MUX system of \cite{Alvarado09c} with optimized constellation and the BICM-S system with a (punctured) code with $R=3/4$ are also shown.}
    \label{UB_vs_snr_n_2_q_3_AWGN}
\end{center}
\end{figure}

In Fig.~\ref{UB_vs_snr_n_2_q_3_AWGN}, we present the results obtained using the proposed system based on the optimum parameters in \eqref{optimal_K_n_2_q_3}--\eqref{optimal_a1_a2_n_2_q_3} and we compare them against four different BICM designs. The first one is the BICM-S system of \cite{Caire98} with an 8-PAM constellation and the second one the BICM system with R-MUX and an 8-PAM constellation of \cite{Alvarado09c}\footnote{The optimum R-MUX for this code is such that 2/3 of the coded bits form the second encoder's output are sent to $k=1$, 1/3 of the bits from the first and the second encoder's outpus are sent to $k=2$, and 2/3 of the bits from the first encoder's output to $k=3$.}. The other two will be explained below. When comparing the proposed system and the BICM-S system of \cite{Caire98}, gains of about 3~dB are observed for a BER target of $10^{-6}$. The gains compared to the system in \cite{Alvarado09c} are about 2.75~dB.

The performance difference between the proposed system and the BICM system with R-MUX and an 8-PAM constellation of \cite{Alvarado09c} are quite large (2.75~dB). However, the comparison is somehow unfair since our system allows HPAM constellations while the results for the system in \cite{Alvarado09c} are given for an 8-PAM constellation. To make a more fair comparison, we have optimized the constellation for the system with R-MUX of \cite{Alvarado09c}, \ie we selected the optimum constellation for each average SNR\footnote{We obtained $\alpha_1(\overline{\gamma})= 0.49$ for all $\overline{\gamma}~[\tr{dB}]\in\set{10, 11,\ld,16}$.}. The obtained results are shown in Fig.~\ref{UB_vs_snr_n_2_q_3_AWGN}. The performance of the system in \cite{Alvarado09c} improves, however, the proposed system still outperforms the R-MUX HPAM in 1.7~dB at a BER of $10^{-6}$. This can be explained by the fact that in the proposed system exploits the temporal behavior of the coded sequence in a deterministic way (via the D-MUX), contrary to the R-MUX of \cite{Alvarado09c}, which assigns the coded bits to different interleavers in a pseudo-random fashion.

Finally, and motivated by the results in \eqref{optimal_K_n_2_q_3}--\eqref{optimal_a1_a2_n_2_q_3}, we analyze a BICM-S system, a 4-PAM constellation, and a puncturing defined by $\mb{P}$ that gives a rate $R=3/4$. This configuration is simply another way of obtaining an spectral efficiency of 1.5~bit/dimension using the same encoder and decoder. We performed an exhaustive search over puncturing patterns and found that for the SNR of interest, the optimal puncturing pattern is the one given by $\mb{P}$. The results are shown in Fig.~\ref{UB_vs_snr_n_2_q_3_AWGN}. In this case, and for a target BER of $10^{-6}$, the proposed system still offers gains of about 0.4~dB.

Now we turn our attention to Nakagami-$m$ fading channels. Since the UB in Theorem~\ref{theo_UB_SPA_Fading} depends on $\mb{K}$, $(\alpha_1(\overline{\gamma}),\alpha_2(\overline{\gamma}))$, $m$, and $\overline{\gamma}$, in general, the optimization must be done jointly over all these parameters. However, we have observed that for a given value of $m$, the optimal constellation and D-MUX do not change significantly for the SNR range of interest. Motivated by this observation, we have found the optimal constellation for an average SNR that gives a BER of approximately $10^{-7}$, and we have used these values for all the range of average SNR. The obtained values are
\begin{align}\label{optKfad}
\mb{K}^*_{m=1}=\mb{K}^*_{m=2}	 =\left[\begin{array}{@{}c@{~}c@{~}c@{}}(2,1)&(3,1)&(3,2)\\(2,2)&(1,1)&(1,2)\end{array}\right],\quad
\mb{K}^*_{m=5}				 =\left[\begin{array}{@{}c@{~}c@{~}c@{}}(1,1)&(2,1)&(3,1)\\(3,2)&(2,2)&(1,2)\end{array}\right],
\end{align}
and
\begin{align}
(\alpha_1^*(\overline{\gamma}),\alpha_2^*(\overline{\gamma}))|_{m=1}&=(0.48, 0.20)\\
(\alpha_1^*(\overline{\gamma}),\alpha_2^*(\overline{\gamma}))|_{m=2}&=(0.47, 0.17)\\
(\alpha_1^*(\overline{\gamma}),\alpha_2^*(\overline{\gamma}))|_{m=5}&=(0.42, 0.01)\label{opta1a2fad}.
\end{align}

We note that by selecting one set of parameters for the range of  average SNR and a  given $m$ is relevant from a practical point of view. This is simply because in practice it would be more difficult to change the constellation parameters and the MUX for each value of $\overline{\gamma}$.

\begin{figure}[t!]
\psfrag{ylabel}[ct][ct][0.8]{$\tr{BER}$}%
\psfrag{xlabel}[cB][cB][0.8]{$\overline{\gamma}$~[dB]}%
\psfrag{BICM-S-ES-QAM}[lc][lc][0.8]{Traditional BICM-S (ES-QAM) \cite{Caire98}}%
\psfrag{R-MUX-optimum-ES-QAM}[lc][lc][0.8]{R-MUX (ES-QAM) \cite{Alvarado09c}}%
\psfrag{S-MUX-optimum-NES-QAM}[lc][lc][0.8]{Optimal HQAM-BICM}%
\psfrag{Sim-BICM-S}[lc][lc][0.8]{Sim. Traditional BICM-S (ES-QAM)}%
\psfrag{Sim-RMUX}[lc][lc][0.8]{Sim. R-MUX (ES-QAM)}%
\psfrag{Sim-SMUX}[lc][lc][0.8]{Sim. Optimal HQAM-BICM}%
\psfrag{m-1}[lc][lc][0.8]{$m=1$}%
\psfrag{m-2}[lc][lc][0.8]{$m=2$}%
\psfrag{m-5}[lc][lc][0.8]{$m=5$}%
\begin{center}
 \includegraphics[width=0.9\columnwidth]{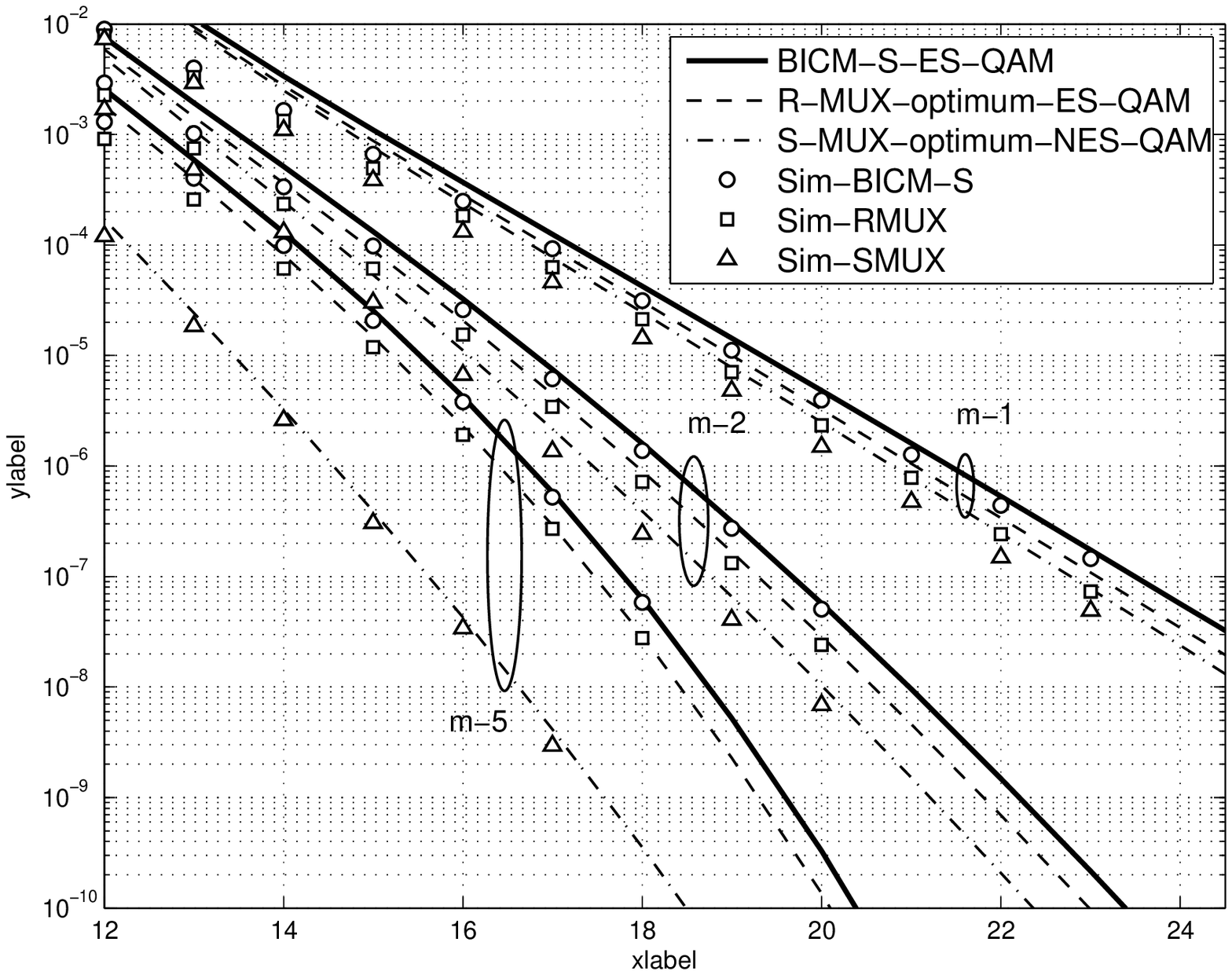}
 \caption{UB (lines) for $n=2$ and $q=3$ (1.5~bit/dimension) for Nakagami-$m$ fading channels given by Theorem~\ref{theo_UB_SPA_Fading}  for $m=1,2,5$. Numerical simulations are also included (markers). The simulations for the proposed HQAM-BICM system uses the values in \eqref{optKfad}--\eqref{opta1a2fad}. The BICM-S system of \cite{Caire98} with an 8-PAM constellation and the BICM system with R-MUX and an 8-PAM constellation of \cite{Alvarado09c} are shown for comparison.}
  \label{BERvsSNR:Example2:Fading}
\end{center}
\end{figure}

In Fig.~\ref{BERvsSNR:Example2:Fading} we present the simulated BER obtained by HQAM-BICM using \eqref{optKfad}--\eqref{opta1a2fad} and the UB with optimized parameters for each SNR. We compare the results obtained by the proposed system against two previous BICM designs (as in Fig.~\ref{BERvsSNR_fading_opt_alpha_opt_MUX_m_1_m_5_m_inf_v2_with_sim}): the BICM-S system of \cite{Caire98} with an 8-PAM constellation and the BICM system with an R-MUX in \cite{Alvarado09c}. From this figure we observe that in fading channels, the proposed HQAM-BICM again outperforms previous BICM designs. When compared to BICM-S, the proposed system offers gains up to 2~dB for $m=5$ and a BER target of $10^{-7}$. The gains compared to the configuration in \cite{Alvarado09c} are less than when compared to BICM-S, but still quite large. This figure also shows that the achievable gains increase when the fading is less severe ($m$ increases). 

\section{Conclusions}\label{Sec:Conclusion}
In this paper we proposed and studied a new BICM transmission framework that uses HQAM constellations in conjunction with a dterministic bit-level multiplexer and M-interleavers. It was shown that a number of degrees of freedom can be exploited, which in turn gives performance improvements of a few decibels compared to previous BICM designs. The gains were shown to depend on the fading parameter, the BER target, and the spectral efficiency, and in general, they increase when the fading is less severe. %It was also shown that even when the general problem is a multidimensional optimization %problem, a solution for a given average SNR gives a good performance for all the range of BER %of interest.

There are a number of degrees of freedom that can be exploited in BICM transmission which may improve its performance even further. In particular, in this paper we only studied HQAM constellations labeled by the BRGC. The performance of BICM with other binary labelings and fully asymmetric constellation is still unknown. Moreover, the period of the MUX gives another degree of freedom not fully exploited in this paper (only short periods were considered).

\appendices

% theo_PEP_SPA
\section{Proof of Theorem~\ref{theo_PEP_SPA}}\label{Appendix.theo_PEP_SPA}

The Laplace transform of the PDF of the decision variable $D(\mb{w}) $ in \eqref{D_definition} is
\begin{equation}
\Phi_{D(\mb{w})}(s;\overline{\gamma}) 	 = \prod_{k=1}^{q}\left[\Phi_{L_k}(s;\overline{\gamma}) \right]^{w_k},
\label{LT:Dec:Var}
\end{equation}
and its cumulant transform $\kappa_{D(\mb{w})}(s;\overline{\gamma})$ is
 \begin{align}
\kappa_{D(\mb{w})}(s;\overline{\gamma}) & = \log[\Phi_{D(\mb{w})}(s;\overline{\gamma})]= \sum_{k=1}^{q}w_k \log \left[\Phi_{L_k}(s;\overline{\gamma}) \right].
\label{Cum:Gen:Fun}
\end{align}

The  PEP can be approximated using the saddlepoint approximation \cite{Martinez06}  as
\begin{align}
\tr{PEP}(\mb{w};\overline{\gamma}) & \approx \frac{1} {\hat{s}\sqrt{2 \pi \kappa^{\prime\prime}_
{D(\mb{w})}(\hat{s};\overline{\gamma})}} \exp(\kappa_{D(\mb{w})}(\hat{s};\overline{\gamma})) 
\label{SPA:1},
\end{align}
where $ \kappa^{\prime\prime}_
{D(\mb{w})}(\hat{s};\overline{\gamma})$ is the second derivative of the cummulant generating function evaluated at the  saddlepoint $\hat{s}$ and can be expressed as
\begin{align}
\kappa_{D(\mb{w})}^{\prime\prime}(\hat{s};\overline{\gamma})  &= \sum_{k=1}^{q}  w_k \left[ \frac{\Phi_{L_k}^{\prime\prime}(\hat{s};\overline{\gamma})  }{\Phi_{L_k}(\hat{s};\overline{\gamma})}\right].
\label{Sec:der:kappa}
\end{align}

Using \eqref{Cum:Gen:Fun} and  \eqref{Sec:der:kappa}  in \eqref{SPA:1} completes the proof.

%Appendix B
\section{Proof of Theorem~\ref{theo_UB_SPA_NonFading}} \label{Appendix.theo_UB_SPA_NonFading}
The two-sided Laplace transform of $p_{L_k}(\lambda;\gamma)$ in \eqref{Gaussain:PDF:mixture}, can be written as 
  \begin{align}
\label{Sec:IntCodeDesign:LaplaceTrans1}
	\Phi_{L_k}(s;\gamma) &= \int_{-\infty}^{\infty} \tr{e}^{-s\lambda} \sum_{j=0}^{M_k-1}\xi_{k,j}\psi(\lambda;\mu_{k,j}\gamma,2|\mu_{k,j}|\gamma)\,\tr{d}\lambda\\
\label{Sec:IntCodeDesign:LaplaceTrans2}
							&=\sum_{j=0}^{M_k-1}\xi_{k,j}\exp\left(\mu_{k,j}\gamma (s^2-s)\right),
							\end{align}
where to pass from \eqref{Sec:IntCodeDesign:LaplaceTrans1}
to \eqref{Sec:IntCodeDesign:LaplaceTrans2} we used the transform pair $\mc{N}(\lambda;\mu,2\mu) \Leftrightarrow
\tr{exp}(\mu(s^2-s))$. The saddlepoint can be found by solving $\Phi_{L_k}'(s;\gamma)=0$, which from \eqref{Sec:IntCodeDesign:LaplaceTrans2} gives $\hat{s}=1/2$.

From \eqref{Sec:IntCodeDesign:LaplaceTrans2}, the second derivative of $\Phi_{L_k}(s;\gamma)$ with respect to $s$ at the saddlepoint ($\hat{s}=1/2$) is
\begin{equation}
\Phi_{L_k}^{\prime\prime}(1/2;\gamma) =\sum_{j=0}^{M_k-1}2\xi_{k,j}\mu_{k,j} \gamma\exp\left(-\mu_{k,j}\gamma/4 \right).
\label{DoubleDerLTNonFading}
							\end{equation}

Substituting \eqref{Sec:IntCodeDesign:LaplaceTrans2} and \eqref{DoubleDerLTNonFading} in \eqref{SPA:final}, the  PEP over the AWGN channel is
\begin{align}
\tr{PEP}(\mb{w};\gamma)&=\left[ \pi \gamma \sum_{k=1}^{q}  w_k \frac{\sum_{j=0}^{M_k-1}\xi_{k,j} \mu_{k,j} \exp\left(-\mu_{k,j}{\gamma}/4\right) }{\sum_{j=0}^{M_k-1}\xi_{k,j}\exp\left(-\mu_{k,j}{\gamma}/4\right)}\right]^{-1/2} \cd  \nonumber\\ &\qquad \qquad \qquad \qquad \qquad \qquad \prod_{k=1}^{q}\left[\sum_{j=0}^{M_k-1}\xi_{k,j}\exp\left(-\mu_{k,j}{\gamma}/4\right)\right]^{w_k} \label{PEP:SPA:AWGN}.
\end{align}
Using \eqref{PEP:SPA:AWGN} in \eqref{Sec:IntCodeDesign:UB_Ndim} completes the proof.

% Appendix
\section{Proof of Theorem~\ref{theo_UB_SPA_Fading}}\label{Appendix.theo_UB_SPA_Fading}
Due to the linearity property of the Laplace transform, we have
\begin{equation}
\Phi_{L_k}(s;\overline{\gamma})=\int_{0}^{\infty} p_{\Gamma}(\gamma;\overline{\gamma}) \Phi_{L_k}(s;\gamma) \,\tr{d}\gamma.
\label{Sec:IntCodeDesign:avgLT1}
\end{equation}
Using \eqref{gamma:dis} and \eqref{Sec:IntCodeDesign:LaplaceTrans2} in \eqref{Sec:IntCodeDesign:avgLT1}, we obtain
\begin{equation}
\Phi_{L_k}(s;\overline{\gamma})=\sum_{j=0}^{M_k-1}\xi_{k,j}\left( \frac{m}{m-\overline{\gamma}\mu_{k,j}(s^2-s)}\right)^m.
\label{Sec:IntCodeDesign:avgLT2}
\end{equation}
From \eqref{Sec:IntCodeDesign:avgLT2}, the saddlepoint is $\hat{s}=1/2$, and the second derivative of $\Phi_{L_k}(s;\overline{\gamma})$ with respect to $s$ at saddlepoint can be written as 
\begin{equation}
\Phi_{L_k}^{\prime\prime}(1/2;\overline{\gamma}) 	 =\sum_{j=0}^{M_k-1}2\xi_{k,j}  \overline{\gamma} \mu_{k,j} \left( \frac{4m}{4m+\overline{\gamma}\mu_{k,j}}\right)^{(m+1)}.
\label{Sec:Der:LT}
\end{equation}
Substituting \eqref{Sec:IntCodeDesign:avgLT2} and \eqref{Sec:Der:LT} in \eqref{SPA:final}, the  PEP over Nakagami-$m$ fading is
\begin{align}
\tr{PEP}(\mb{w};\overline{\gamma})&= \left[\pi  \overline{\gamma}  \sum_{k=1}^{q}  w_k \frac{\sum_{j=0}^{M_k-1}\xi_{k,j} \mu_{k,j} \left( \frac{4m}{4m+\overline{\gamma}\mu_{k,j}}\right)^{(m+1)} }{\sum_{j=0}^{M_k-1}\xi_{k,j}\left( \frac{4m}{4m+\overline{\gamma}\mu_{k,j}}\right)^m}\right]^{-1/2} \prod_{k=1}^{q}\left[\sum_{j=0}^{M_k-1}\xi_{k,j}\left( \frac{4m}{4m+\overline{\gamma}\mu_{k,j}}\right)^m\right]^{w_k}.
\label{PEP_SPA_Fading}
\end{align}
Using  \eqref{PEP_SPA_Fading} in \eqref{Sec:IntCodeDesign:UB_Ndim} completes the proof.

\bibliography{IEEEabrv,references_all}
\bibliographystyle{IEEEtran}

\end{document}